\documentclass{elsart}

\usepackage{graphicx}
\usepackage{amssymb}
\usepackage{sparticles}
\usepackage{feynmf}

\begin{document}

\def\psibar{\overline\psi}
\def\tr{\mathop{\rm tr}\nolimits}
\def\Tr{\mathop{\rm Tr}\nolimits}
\def\zz{\hphantom{-}}          %% use in math mode
\def\xx{\hphantom{0}}          %% use in math mode
\def\gsim{\lower0.6ex\vbox{\hbox{$\ \buildrel{\textstyle >}\over{\sim}\ $}}} 
\def\lsim{\lower0.6ex\vbox{\hbox{$\ \buildrel{\textstyle <}\over{\sim}\ $}}} 

\begin{frontmatter}



\title{Three-loop Corrections in a Covariant Effective Field Theory}


\author{Jeff McIntire}
\ead{oberonjwm@yahoo.com}
\address{Department of Physics and Nuclear Theory Center \\
Indiana University \\
Bloomington, IN\ \ 47405}

\begin{abstract}
Chiral effective field theories have been used with success in the
study of nuclear structure. It is of interest to systematically improve these energy
functionals (particularly that of quantum hadrodynamics) 
through the inclusion of many-body correlations. One possible
source of improvement is the loop expansion. Using the techniques of 
Infrared Regularization, the short-range, local dynamics at each order in the loops
is absorbed into the
parameterization of the underlying effective lagrangian. The remaining nonlocal,
exchange correlations must be calculated explicitly. 
Given that the interactions of quantum hadrodynamics are relatively 
soft, the loop expansion may be manageable or even perturbative in nuclear matter.
This work investigates the role played by the three-loop contributions to the 
loop expansion for quantum hadrodynamics.
\end{abstract}

\end{frontmatter}

% main text

\section{Introduction}

Density Functional Theory (DFT) is a powerful technique originally developed for use in condensed matter physics
\cite{ref:Ho64,ref:Ko65,ref:Ko99} that has been successfully adapted to nuclear physics 
\cite{ref:Dr90,ref:Sp92,ref:Sc95,ref:Fu97,ref:Se97,ref:Wa04}. DFT states that the ground-state expectation 
value of any observable is a {\it unique} functional of the exact ground-state density; moreover, 
if the expectation value of the hamiltonian is considered as a functional of the density, then
the exact ground-state density can be determined by minimizing the energy functional \cite{ref:Ho64,ref:Ko99}.
Furthermore, DFT allows one to replace 
the quantum many-body equations by a series of single-particle equations with local, classical fields that reproduce 
certain observables exactly (energy, scalar and vector densities, and chemical potential) \cite{ref:Ko65,ref:Wa04,ref:Al85}.
Thus, the problem is reduced to determining the exact ground-state energy functional. 
However, this is impossible in practice. As a result, a number
of approximate energy functionals have been developed; one such theory is 
based on quantum hadrodynamics (QHD).

QHD is a low-energy theory of the strong interaction 
\cite{ref:Fu97,ref:Se97,ref:Wa04,ref:Wa74,ref:Fu99,ref:Fu03,ref:Se04,ref:Fu04}. 
Here the hadron, and not the quark, is
the observed degree of freedom (due to confinement at this energy scale). QHD models the nuclear force
as a exchange of mesons between nucleons. Isoscalar scalar ($\sigma$) and vector ($\omega$) mesons 
represent a medium-range attraction and a short-range repulsion respectively. The pion is
also included to take chiral symmetry into account. DFT allows one to replace the quantum meson
fields with their classical equivalents (Kohn-Sham potentials called mean fields); these mean fields, while large, 
are small compared to the chiral symmetry breaking scale. As a result, one
can use these ratios as small parameters with which to expand the energy functional in a controlled fashion.
Each term in this lagrangian is characterized by an 
undetermined coefficient which is assumed to be natural, or order unity \cite{ref:Ma84,ref:Ge93}. 
The resulting lagrangian \cite{ref:Fu97,ref:Se97} has provided a method for predicting the properties of nuclei 
\cite{ref:Mc02,ref:Mc04,ref:Mc05,ref:He02,ref:He03,ref:He04}.

However, the mean field theory of QHD is only an approximation to the exact energy functional. The question remains how this theory
behaves when other many-body corrections are included to improve the energy functional 
(such as loops, rings, clustering, etc.). This work
is part of an investigation into the effects of loops on this particular energy functional and is a continuation
of \cite{ref:Mc07,ref:Mc07a}. In these works, the many-body loop expansion was carried out to the two-loop order where,
using the techniques of Infrared Regularization, all of the short-range, local dynamics were absorbed into
the parameterization of the underlying effective lagrangian. What remained was the nonlocal exchange 
correlations, which were then explicitly calculated. In addition, the effect of this expansion
on naturalness was investigated and the size of the two-loop exchange integrals was determined. Since the interactions
of QHD are relatively soft, one might expect that, once Infrared Regularization is taken into account,
the loop expansion may be asymptotic. It is the purpose of this work to investigate the 
effect of the three-loop contributions on this theory.

The loop expansion for QHD is constructed in the usual manner 
\cite{ref:Fu89,ref:Di33,ref:It80,ref:Se86,ref:Co73,ref:Il75,ref:Co77}. 
The effective action is expanded around its classical value by grouping terms according to the 
number of quantum loops (or powers of $\hbar$) in their corresponding diagrams. 
For the purposes of this work, we are interested only in the terms in the expansion
at the three-loop level. All of the integrals that represent tadpole and disconnected diagrams cancel out 
in the effective action and we are left with only the fully connected diagrams. 
In addition, those diagrams which are anomalous are discarded.
For the cases considered here, there are eleven integrals of interest.
Ten of these integrals have four factors of the nucleon propagator and two meson 
propagators (either scalar, vector, or pion) each. There is one additional three-loop diagram with
two factors of both the baryon and the pion propagator. In this work,
no nonlinearities in the isoscalar mesons were included in the effective lagrangian, as in \cite{ref:Mc07}.
In \cite{ref:Mc07a}, the effect of these nonlinearities was explored at the two-loop level.
As we are interested only in the general effects of the loop expansion at third order
(and not an improved equation of state), these nonlinearities are not retained. The effect of the inclusion
of nonlinear meson self-interactions on the three-loop integrals is left for future work.

Why a loop expansion? The loop expansion is a simple and well-developed expansion scheme in powers
of $\hbar$ that is derived from the path integral. The mean meson fields are included non-perturbatively
and the correlations are included perturbatively. Therefore, one can analyze the many-body effects
order by order. The loop expansion has the advantage that it is fairly easy to separate the local and 
nonlocal dynamics and analyze their structures \cite{ref:Mc07,ref:Mc07a,ref:Ta96,ref:El98,ref:Be99,ref:Be00,ref:Sc03}. 
Consideration of the three-loop integrals will help
us determine if this is a good expansion scheme or not for QHD. 
 
The diagrams at the three-loop level can be written as either ladders or crossed ladders
(with the exception of the one additional pion diagram). If one considers the Bethe-Saltpeter
equation in the ladder approximation, one can replace the two nucleon propagators in the ladder with an approximate two-body
propagator \cite{ref:Se86,ref:Br76,ref:Ho87,ref:Bl66,ref:Th70,ref:Er72}. 
This two-body propagator contains the nonlocal physics from the two one-body propagators
and provides the proper cuts to maintain analyticity. It describes the propagation of two nucleons above the
Fermi surface. As there is some ambiguity in the definition of this propagator, we will consider three
different versions: Blankenbecler-Sugar, Thompson, and Erkelenz-Holinde \cite{ref:Se86,ref:Bl66,ref:Th70,ref:Er72}. 

The remaining pion (football) diagram arises
from the inclusion of a two-pion vertex. Isospin and parity considerations restrict the two pion
vertex to diagrams with the pions and nucleons only, at least at the three-loop level. 
Because this vertex is antisymmetric, only a diagram that has two of these vertices will survive.
Fortunately, this diagram is analogous to the ladders and crossed ladders; here we use
a two-body pion propagator (essentially the same as the baryon case but without the projection operators
and theta functions) \cite{ref:Br76}.
Note that this two-pion exchange is not equivalent to scalar ($\sigma$) exchange (which is
an effective field representing any interaction in the isoscalar-scalar channel).
Any potential redundancy is eliminated by the parameterization.

The remaining nucleon propagators can be separated into two components: Feynman and Density
parts \cite{ref:Fu89,ref:Se86}. This is accomplished by taking into account the proper pole structure of the propagator.
The Feynman part describes the propagation of a baryon or an antibaryon; the Density portion
describes on-shell propagation of a nucleon while taking the exclusion principle into account.
Thus, one can separate the local and nonlocal dynamics, as in \cite{ref:Mc07,ref:Mc07a}
(see also Infrared Regularization discussions in \cite{ref:Ta96,ref:El98,ref:Be99,ref:Be00,ref:Sc03}).
All of the vacuum loops are parameterized; only the boson exchange between valence nucleons is 
calculated.

It is also of interest to investigate how these loop integrals fit into the power counting
scheme that was developed for the underlying QHD lagrangian \linebreak \cite{ref:Fu97,ref:Se97,ref:Se04,ref:Fu04}. 
It was shown in previous work that the two-loop integrals are about third order in the power counting \cite{ref:Mc07,ref:Mc07a}. 
In this work, we will determine the size of the three-loop integrals and how they relate
to both the one- and two-loop levels. The tricky business of finding the appropriate expansion parameter
is left for future work.

The loop expansion has been studied in the context of other effective field theories, particularly
Chiral Perturbation Theory (ChPT) \cite{ref:Ta96,ref:El98,ref:Be99,ref:Be00,ref:Sc03,ref:Ka01,ref:Vr04}.
The techniques of Infrared Regularization were developed to separate out and absorb the local portions
of the loop expansion in ChPT \cite{ref:Ta96,ref:El98,ref:Be99,ref:Be00,ref:Sc03}. More recently,
investigations have been conducted to determine the effect of multi-loop correlations in ChPT and to
discover the appropriate expansion parameter \cite{ref:Ka01,ref:Vr04}.

In this paper, the many-body loop expansion for QHD is investigated at the three-loop level.
Infrared Regularization allows us to parameterize the local dynamics. The nonlocal exchange correlations are solved in the ladder
approximation to the Bethe-Saltpeter equation for the lowest-order truncation in the
underlying lagrangian. The effects of a high momentum cutoff are explored. The naturalness of the lowest level couplings
is confirmed at the three-loop level. In addition, the size of the integrals is discussed and how they fit into
the power counting scheme is investigated.

\section{Theory}
\label{sec:1}

In this section, we present the corrections to QHD arising from the three-loop contributions to the loop expansion.
The background for the loop expansion in QHD is presented in previous works \cite{ref:Mc07,ref:Mc07a,ref:Fu89}.
The lagrangian used here follows from the {\it chirally invariant\/} lagrangian in \cite{ref:Fu97,ref:Se97}, or
\begin{eqnarray} {\cal L} & = &
-{\psibar}\left[\gamma_{\mu}\left(\partial_{\mu} -
ig_{V}V_{\mu}\right) - ig_{A}\gamma_{\mu}\gamma_{5}a_{\mu} + i\gamma_{\mu}v_{\mu}
+ \left(M-g_{S}\phi\right)\right]\psi \nonumber \\
& & - \frac{1}{2}\left(\partial_{\mu}\pi_{a}\right)^{2} - \frac{1}{2}m_{\pi}^{2}\pi_{a}^{2}
 \ , \label{eqn:lagrangian}
\end{eqnarray}
where $\underline{\pi}=\frac{1}{2}\pi_{a}\cdot\tau_{a}$. Here $\psi$ are
the fermion fields and $\phi$, $V_{\mu}$, and $\pi_{a}$ are the
meson fields (isoscalar-scalar, isoscalar-vector, and
isovector-pseudoscalar, respectively).
The heavy meson fields are also chiral scalars.
Note that in this work, the conventions of \cite{ref:Wa04} are used.
As in \cite{ref:Mc07,ref:Mc07a}, we are not making a chiral expansion in powers of the pion mass and
include it for kinematical purposes only. The pion-nucleon interactions
retained above can be expanded in the following fashion (with $\underline{\xi}=\exp\{-i\underline{\pi}/f_{\pi}\}$) 
\cite{ref:Fu97,ref:Se97}:
\begin{eqnarray}
v_{\mu} & = & -\frac{i}{2}\left(\underline{\xi}^{\dagger}\partial_{\mu}\underline{\xi}
+\underline{\xi}\partial_{\mu}\underline{\xi}^{\dagger}\right) 
= -\frac{i}{2f_{\pi}^{2}}\left[\underline{\pi},\partial_{\mu}\underline{\pi}\right] +\cdots \ , \\
a_{\mu} & = & -\frac{i}{2}\left(\underline{\xi}^{\dagger}\partial_{\mu}\underline{\xi}
-\underline{\xi}\partial_{\mu}\underline{\xi}^{\dagger}\right)
= -\frac{1}{f_{\pi}}\partial_{\mu}\underline{\pi} +\cdots \ .
\end{eqnarray}
\noindent It is necessary to retain only the first term in $v_{\mu}$ and $a_{\mu}$ at the three-loop level.
Note that the first term in $v_{\mu}$ is a two-pion--nucleon vertex that does not contribute at the two-loop level. 

The generating functional is defined in the usual way by
\begin{eqnarray}
Z[j,J_{\mu}] & \equiv & \exp\left\{iW[j,J_{\mu}]/\hbar\right\} \nonumber \\
& = & {\cal N}^{-1}\int D({\psibar})D(\psi)D(\phi)D(V_{\mu})D(\pi_{a}) \nonumber \\
& & \times \exp\left\{\frac{i}{\hbar} \int d^{4}x\left[{\cal L}(x) +
j(x)\phi(x) + J_{\mu}(x)V_{\mu}(x)\right]\right\}\ , \label{eqn:3}
\end{eqnarray}
\noindent where $\cal N$ is the normalization factor (in effect, the
vacuum subtraction), $j(x)$ and $J_{\mu}(x)$ are the external
sources corresponding to the meson fields $\phi$ and $V_{\mu}$,
respectively, and the connected generating functional is $W[j,J_{\mu}]$.

\subsection{Ladders and Crossed Ladders}
\label{sec:1a}

Consider the portion of the connected generating functional that describes the scalar-scalar three-loop 
contributions, which is
\begin{eqnarray}
W_{3-SS} & = & -i\frac{\hbar^{3}}{24}g_{S}^{4} \int\int d^{4}xd^{4}yd^{4}zd^{4}a \nonumber \\
& & \times \left[\frac{-i\delta}{\delta u(x)}\right]\left[\frac{-i\delta}{\delta u(y)}\right]
\left[\frac{-i\delta}{\delta u(z)}\right]\left[\frac{-i\delta}{\delta u(a)}\right] \nonumber \\
& & \times \left[\frac{i\delta}{\delta \xi(x)}\right]_{\alpha}\left[\frac{-i\delta}{\delta \bar{\xi}(x)}\right]_{\alpha'}
\left[\frac{i\delta}{\delta \xi(y)}\right]_{\beta}\left[\frac{-i\delta}{\delta \bar{\xi}(y)}\right]_{\beta'} \nonumber \\ 
& & \times \left[\frac{i\delta}{\delta \xi(z)}\right]_{\gamma}\left[\frac{-i\delta}{\delta \bar{\xi}(z)}\right]_{\gamma'}
\left[\frac{i\delta}{\delta \xi(a)}\right]_{\mu}\left[\frac{-i\delta}{\delta \bar{\xi}(a)}\right]_{\mu'} \nonumber \\ 
& & \times \exp\left\{-i \int\int d^{4}x_{1}d^{4}x_{2} \bar{\xi}(x_{1})G_{H}(x_{1}-x_{2})\xi(x_{2})\right\} \nonumber \\ & & 
\left. \times \exp\left\{\frac{i}{2}\int\int d^{4}x_{1}d^{4}x_{2}u(x_{1})\Delta_{S}^{0}(x_{1}-x_{2})u(x_{2})\right\}\right|_{sources=0}
\nonumber \\ & & - VEV \ .
\end{eqnarray}

\noindent We then perform the variational derivatives and drop the terms with the 
tadpole diagrams (these diagrams will cancel when we consider the effective action). Three of the remaining terms are 
disconnected diagrams (the ones that are essentially the scalar two-loop integral squared); these are discarded for the
same reason. Next, we perform Fourier transforms over all the terms. The three delta functions that arise let us 
eliminate three of the six momentum integrals. Rearranging and combining terms, we get 
(after transforming to the energy density and suppressing the $\hbar$)
\begin{eqnarray}
{\cal E}_{3-SS} & = & -i\frac{g_{S}^{4}}{4}\int\int\frac{d^{4}k}{(2\pi)^{4}}
\frac{d^{4}q}{(2\pi)^{4}}\frac{d^{4}q'}{(2\pi)^{4}}  \nonumber \\
& & \times \left\{\Delta_{S}(k)\Delta_{S}(k)\tr\left[G_{H}(q-k)G_{H}(q)\right]\tr\left[G_{H}(q')G_{H}(q'+k)\right]\right. \nonumber \\
& & -\Delta_{S}(k)\Delta_{S}(q'-k)\tr\left[G_{H}(q'+q-k)G_{H}(q)G_{H}(q+k)G_{H}(q'+q)\right] \nonumber \\
& & \left. -2\Delta_{S}(k)\Delta_{S}(q-q')\tr\left[G_{H}(q-k)G_{H}(q)G_{H}(q')G_{H}(q)\right]\right\} \ . \nonumber \\ & &
\label{eqn:tot}
\end{eqnarray}

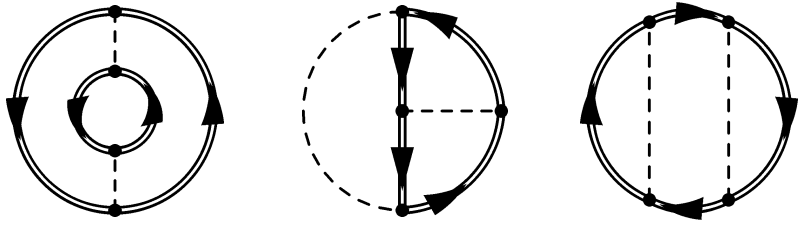
\begin{figure}[!htb]
  \begin{center}
    \begin{tabular}{ccc}
      \begin{fmffile}{ss2}
	\fmfframe(1,2)(1,2){ 
	  \begin{fmfgraph}(75,75)
	    \fmfipair{a,b,c,d,e,f,g,h,i,j}
	    \fmfiequ{a}{.5[nw,ne]}
	    \fmfiequ{b}{.5[sw,se]}
	    \fmfiequ{c}{.7[a,b]}
	    \fmfiequ{d}{.3[a,b]}
	    \fmfiequ{e}{.3[nw,ne]}
	    \fmfiequ{f}{.3[sw,se]}
	    \fmfiequ{g}{.5[e,f]}
	    \fmfiequ{h}{.7[nw,ne]}
	    \fmfiequ{i}{.7[sw,se]}
	    \fmfiequ{j}{.5[h,i]}
	    \fmfi{dbl_plain_arrow}{a .. .5[nw,sw] .. .b}
	    \fmfi{dbl_plain_arrow}{b .. .5[ne,se] .. .a}
	    \fmfi{dbl_plain_arrow}{d .. g .. c}
	    \fmfi{dbl_plain_arrow}{c .. j .. d}
	    \fmfi{dashes}{a .. d}
	    \fmfi{dashes}{b .. c}
	    \fmfiv{d.sh=circle,d.siz=2thick}{a}
	    \fmfiv{d.sh=circle,d.siz=2thick}{b}
	    \fmfiv{d.sh=circle,d.siz=2thick}{c}
	    \fmfiv{d.sh=circle,d.siz=2thick}{d}
	  \end{fmfgraph}
	}
      \end{fmffile}

      &

      \begin{fmffile}{ss3}
	\fmfframe(1,2)(1,2){ 
	  \begin{fmfgraph}(75,75)
	    \fmfipair{a,b,c,d,e,f,g,h}
	    \fmfiequ{a}{.5[nw,ne]}
	    \fmfiequ{b}{.5[sw,se]}
	    \fmfiequ{c}{.5[a,b]}
	    \fmfiequ{d}{.5[ne,se]}
	    \fmfiequ{e}{.7[nw,ne]}
	    \fmfiequ{f}{.7[sw,se]}
	    \fmfiequ{g}{.95[e,f]}
	    \fmfiequ{h}{.05[e,f]}
	    \fmfi{dashes}{a .. .5[nw,sw] .. b}
	    \fmfi{dbl_plain_arrow}{b .. g .. .5[ne,se]}
	    \fmfi{dbl_plain_arrow}{.5[ne,se] .. h .. a}
	    \fmfi{dashes}{c .. d}
	    \fmfi{dbl_plain_arrow}{a .. c}
	    \fmfi{dbl_plain_arrow}{c .. b}
	    \fmfiv{d.sh=circle,d.siz=2thick}{a}
	    \fmfiv{d.sh=circle,d.siz=2thick}{b}
	    \fmfiv{d.sh=circle,d.siz=2thick}{c}
	    \fmfiv{d.sh=circle,d.siz=2thick}{d}
	  \end{fmfgraph}
	}
      \end{fmffile}
      
      &
      
      \begin{fmffile}{ss1}
	\fmfframe(1,2)(1,2){ 
	  \begin{fmfgraph}(75,75)
	    \fmfipair{a,b,c,d,e,f,g,h}
	    \fmfiequ{a}{.3[nw,ne]}
	    \fmfiequ{b}{.3[sw,se]}
	    \fmfiequ{c}{.7[nw,ne]}
	    \fmfiequ{d}{.7[sw,se]}
	    \fmfiequ{e}{.95[a,b]}
	    \fmfiequ{f}{.05[a,b]}
	    \fmfiequ{g}{.95[c,d]}
	    \fmfiequ{h}{.05[c,d]}
	    \fmfi{dbl_plain_arrow}{e .. .5[nw,sw] .. f}
	    \fmfi{dbl_plain_arrow}{f .. .5[nw,ne] .. h}
	    \fmfi{dbl_plain_arrow}{h .. .5[ne,se] .. g}
	    \fmfi{dbl_plain_arrow}{g .. .5[sw,se] .. e}
	    \fmfi{dashes}{e .. .f}
	    \fmfi{dashes}{g .. .h}
	    \fmfiv{d.sh=circle,d.siz=2thick}{e}
	    \fmfiv{d.sh=circle,d.siz=2thick}{f}
	    \fmfiv{d.sh=circle,d.siz=2thick}{g}
	    \fmfiv{d.sh=circle,d.siz=2thick}{h}
	  \end{fmfgraph}
	}
      \end{fmffile}
    \end{tabular}
    \caption{Scalar-scalar three-loop diagrams defined in Eq.\ (\ref{eqn:tot}). Here the double line represents
    the baryon propagator and the dashed line represents the scalar meson propagator.}
    \label{fey8}
  \end{center}
\end{figure}

\begin{figure}[!htb]
  \begin{center}
    \begin{tabular}{cc}
      \begin{fmffile}{ss5}
	\fmfframe(1,2)(1,2){ 
	  \begin{fmfgraph}(75,75)
	    \fmfipair{a,b,c,d,e,f,g,h,i,j}
	    \fmfiequ{a}{.5[nw,ne]}
	    \fmfiequ{b}{.5[sw,se]}
	    \fmfiequ{c}{.5[nw,sw]}
	    \fmfiequ{d}{.5[ne,se]}
	    \fmfiequ{e}{.3[nw,ne]}
	    \fmfiequ{f}{.3[sw,se]}
	    \fmfiequ{g}{.7[nw,ne]}
	    \fmfiequ{h}{.7[sw,se]}
	    \fmfiequ{i}{.5[e,f]}
	    \fmfiequ{j}{.5[g,h]}
	    \fmfi{dbl_plain_arrow}{b .. c .. a}
	    \fmfi{dbl_plain_arrow}{b .. d .. a}
	    \fmfi{dbl_plain_arrow}{a .. i .. b}
	    \fmfi{dbl_plain_arrow}{a .. j .. b}
	    \fmfiv{d.sh=circle,d.siz=2thick}{a}
	    \fmfiv{d.sh=circle,d.siz=2thick}{b}
	  \end{fmfgraph}
	}
      \end{fmffile}

      &

      \begin{fmffile}{ss4}
	\fmfframe(1,2)(1,2){ 
	  \begin{fmfgraph}(110,75)
	    \fmfipair{a,b,c,d,e,f,g,h,i,j}
	    \fmfiequ{a}{.3[nw,ne]}
	    \fmfiequ{b}{.3[sw,se]}
	    \fmfiequ{c}{.5[nw,ne]}
	    \fmfiequ{d}{.5[sw,se]}
	    \fmfiequ{e}{.7[nw,ne]}
	    \fmfiequ{f}{.7[sw,se]}
	    \fmfiequ{g}{.5[a,b]}
	    \fmfiequ{h}{.5[e,f]}
	    \fmfiequ{i}{.75[c,d]}
	    \fmfiequ{j}{.25[c,d]}
	    \fmfi{dbl_plain_arrow}{g .. i .. h}
	    \fmfi{dbl_plain_arrow}{h .. j .. g}
	    \fmfi{dbl_plain_arrow}{h .. .5[ne,se] .. h}
	    \fmfi{dbl_plain_arrow}{g .. .5[nw,sw] .. g}
	    \fmfiv{d.sh=circle,d.siz=2thick}{g}
	    \fmfiv{d.sh=circle,d.siz=2thick}{h}
	  \end{fmfgraph}
	}
      \end{fmffile}
    \end{tabular}
    \caption{Scalar-scalar three-loop diagrams in which the meson interactions were shrunken to a point. The first and second
    diagrams in Fig.\ \ref{fey8} appear the same here although there are two ways to do the trace(s).}
    \label{fey9}
  \end{center}
\end{figure}
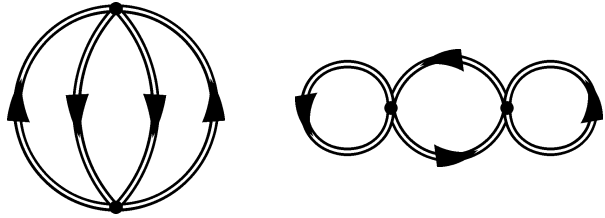

\begin{figure}[!htb]
  \begin{center}
    \begin{tabular}{cccc}
      \begin{fmffile}{ss6}
	\fmfframe(1,2)(1,2){ 
	  \begin{fmfgraph}(75,75)
	    \fmfipair{a,b,c,d,e,f,g,h,i,j}
	    \fmfiequ{a}{.15[nw,ne]}
	    \fmfiequ{b}{.15[sw,se]}
	    \fmfiequ{c}{.85[nw,ne]}
	    \fmfiequ{d}{.85[sw,se]}
	    \fmfiequ{e}{.3[nw,ne]}
	    \fmfiequ{f}{.3[sw,se]}
	    \fmfiequ{g}{.5[e,f]}
	    \fmfiequ{h}{.7[nw,ne]}
	    \fmfiequ{i}{.7[sw,se]}
	    \fmfiequ{j}{.5[h,i]}
	    \fmfi{dbl_plain_arrow}{a .. .5[nw,sw] .. b}
	    \fmfi{dbl_plain_arrow}{b .. g .. a}
	    \fmfi{dbl_plain_arrow}{c .. .5[ne,se] .. d}
	    \fmfi{dbl_plain_arrow}{d .. j .. c}
	    \fmfi{dashes}{a .. .c}
	    \fmfi{dashes}{b .. .d}
	    \fmfiv{d.sh=circle,d.siz=2thick}{a}
	    \fmfiv{d.sh=circle,d.siz=2thick}{b}
	    \fmfiv{d.sh=circle,d.siz=2thick}{c}
	    \fmfiv{d.sh=circle,d.siz=2thick}{d}
	  \end{fmfgraph}
	}
      \end{fmffile}

      &

      \begin{fmffile}{ss7}
	\fmfframe(1,2)(1,2){ 
	  \begin{fmfgraph}(75,75)
	    \fmfipair{a,b,c,d,e,f}
	    \fmfiequ{a}{.15[nw,ne]}
	    \fmfiequ{b}{.15[sw,se]}
	    \fmfiequ{c}{.85[nw,ne]}
	    \fmfiequ{d}{.85[sw,se]}
	    \fmfiequ{e}{.5[nw,ne]}
	    \fmfiequ{f}{.5[sw,se]}
	    \fmfi{dbl_plain_arrow}{a .. .5[nw,sw] .. b}
	    \fmfi{dbl_plain}{d .. a}
	    \fmfi{dbl_plain_arrow}{d .. .5[e,f]}
	    \fmfi{dbl_plain_arrow}{c .. .5[ne,se] .. d}
	    \fmfi{dbl_plain}{b .. c}
	    \fmfi{dbl_plain_arrow}{b .. .5[e,f]}
	    \fmfi{dashes}{a .. .c}
	    \fmfi{dashes}{b .. .d}
	    \fmfiv{d.sh=circle,d.siz=2thick}{a}
	    \fmfiv{d.sh=circle,d.siz=2thick}{b}
	    \fmfiv{d.sh=circle,d.siz=2thick}{c}
	    \fmfiv{d.sh=circle,d.siz=2thick}{d}
	  \end{fmfgraph}
	}
      \end{fmffile}

      &

      \begin{fmffile}{sv6}
	\fmfframe(1,2)(1,2){ 
	  \begin{fmfgraph}(75,75)
	    \fmfipair{a,b,c,d,e,f,g,h,i,j}
	    \fmfiequ{a}{.15[nw,ne]}
	    \fmfiequ{b}{.15[sw,se]}
	    \fmfiequ{c}{.85[nw,ne]}
	    \fmfiequ{d}{.85[sw,se]}
	    \fmfiequ{e}{.3[nw,ne]}
	    \fmfiequ{f}{.3[sw,se]}
	    \fmfiequ{g}{.5[e,f]}
	    \fmfiequ{h}{.7[nw,ne]}
	    \fmfiequ{i}{.7[sw,se]}
	    \fmfiequ{j}{.5[h,i]}
	    \fmfi{dbl_plain_arrow}{a .. .5[nw,sw] .. b}
	    \fmfi{dbl_plain_arrow}{b .. g .. a}
	    \fmfi{dbl_plain_arrow}{c .. .5[ne,se] .. d}
	    \fmfi{dbl_plain_arrow}{d .. j .. c}
	    \fmfi{dashes}{a .. .c}
	    \fmfi{photon}{b .. .d}
	    \fmfiv{d.sh=circle,d.siz=2thick}{a}
	    \fmfiv{d.sh=circle,d.siz=2thick}{b}
	    \fmfiv{d.sh=circle,d.siz=2thick}{c}
	    \fmfiv{d.sh=circle,d.siz=2thick}{d}
	  \end{fmfgraph}
	}
      \end{fmffile}

      \\
      \\

      \begin{fmffile}{sv7}
	\fmfframe(1,2)(1,2){ 
	  \begin{fmfgraph}(75,75)
	    \fmfipair{a,b,c,d,e,f}
	    \fmfiequ{a}{.15[nw,ne]}
	    \fmfiequ{b}{.15[sw,se]}
	    \fmfiequ{c}{.85[nw,ne]}
	    \fmfiequ{d}{.85[sw,se]}
	    \fmfiequ{e}{.5[nw,ne]}
	    \fmfiequ{f}{.5[sw,se]}
	    \fmfi{dbl_plain_arrow}{a .. .5[nw,sw] .. b}
	    \fmfi{dbl_plain}{d .. a}
	    \fmfi{dbl_plain_arrow}{d .. .5[e,f]}
	    \fmfi{dbl_plain_arrow}{c .. .5[ne,se] .. d}
	    \fmfi{dbl_plain}{b .. c}
	    \fmfi{dbl_plain_arrow}{b .. .5[e,f]}
	    \fmfi{dashes}{a .. .c}
	    \fmfi{photon}{b .. .d}
	    \fmfiv{d.sh=circle,d.siz=2thick}{a}
	    \fmfiv{d.sh=circle,d.siz=2thick}{b}
	    \fmfiv{d.sh=circle,d.siz=2thick}{c}
	    \fmfiv{d.sh=circle,d.siz=2thick}{d}
	  \end{fmfgraph}
	}
      \end{fmffile}

      &

      \begin{fmffile}{vv6}
	\fmfframe(1,2)(1,2){ 
	  \begin{fmfgraph}(75,75)
	    \fmfipair{a,b,c,d,e,f,g,h,i,j}
	    \fmfiequ{a}{.15[nw,ne]}
	    \fmfiequ{b}{.15[sw,se]}
	    \fmfiequ{c}{.85[nw,ne]}
	    \fmfiequ{d}{.85[sw,se]}
	    \fmfiequ{e}{.3[nw,ne]}
	    \fmfiequ{f}{.3[sw,se]}
	    \fmfiequ{g}{.5[e,f]}
	    \fmfiequ{h}{.7[nw,ne]}
	    \fmfiequ{i}{.7[sw,se]}
	    \fmfiequ{j}{.5[h,i]}
	    \fmfi{dbl_plain_arrow}{a .. .5[nw,sw] .. b}
	    \fmfi{dbl_plain_arrow}{b .. g .. a}
	    \fmfi{dbl_plain_arrow}{c .. .5[ne,se] .. d}
	    \fmfi{dbl_plain_arrow}{d .. j .. c}
	    \fmfi{photon}{a .. .c}
	    \fmfi{photon}{b .. .d}
	    \fmfiv{d.sh=circle,d.siz=2thick}{a}
	    \fmfiv{d.sh=circle,d.siz=2thick}{b}
	    \fmfiv{d.sh=circle,d.siz=2thick}{c}
	    \fmfiv{d.sh=circle,d.siz=2thick}{d}
	  \end{fmfgraph}
	}
      \end{fmffile}

      &

      \begin{fmffile}{vv7}
	\fmfframe(1,2)(1,2){ 
	  \begin{fmfgraph}(75,75)
	    \fmfipair{a,b,c,d,e,f}
	    \fmfiequ{a}{.15[nw,ne]}
	    \fmfiequ{b}{.15[sw,se]}
	    \fmfiequ{c}{.85[nw,ne]}
	    \fmfiequ{d}{.85[sw,se]}
	    \fmfiequ{e}{.5[nw,ne]}
	    \fmfiequ{f}{.5[sw,se]}
	    \fmfi{dbl_plain_arrow}{a .. .5[nw,sw] .. b}
	    \fmfi{dbl_plain}{d .. a}
	    \fmfi{dbl_plain_arrow}{d .. .5[e,f]}
	    \fmfi{dbl_plain_arrow}{c .. .5[ne,se] .. d}
	    \fmfi{dbl_plain}{b .. c}
	    \fmfi{dbl_plain_arrow}{b .. .5[e,f]}
	    \fmfi{photon}{a .. .c}
	    \fmfi{photon}{b .. .d}
	    \fmfiv{d.sh=circle,d.siz=2thick}{a}
	    \fmfiv{d.sh=circle,d.siz=2thick}{b}
	    \fmfiv{d.sh=circle,d.siz=2thick}{c}
	    \fmfiv{d.sh=circle,d.siz=2thick}{d}
	  \end{fmfgraph}
	}
      \end{fmffile}

      \\
      \\

      \begin{fmffile}{sp7}
	\fmfframe(1,2)(1,2){ 
	  \begin{fmfgraph}(75,75)
	    \fmfipair{a,b,c,d,e,f}
	    \fmfiequ{a}{.15[nw,ne]}
	    \fmfiequ{b}{.15[sw,se]}
	    \fmfiequ{c}{.85[nw,ne]}
	    \fmfiequ{d}{.85[sw,se]}
	    \fmfiequ{e}{.5[nw,ne]}
	    \fmfiequ{f}{.5[sw,se]}
	    \fmfi{dbl_plain_arrow}{a .. .5[nw,sw] .. b}
	    \fmfi{dbl_plain}{d .. a}
	    \fmfi{dbl_plain_arrow}{d .. .5[e,f]}
	    \fmfi{dbl_plain_arrow}{c .. .5[ne,se] .. d}
	    \fmfi{dbl_plain}{b .. c}
	    \fmfi{dbl_plain_arrow}{b .. .5[e,f]}
	    \fmfi{dashes}{a .. .c}
	    \fmfi{dbl_dots}{b .. .d}
	    \fmfiv{d.sh=circle,d.siz=2thick}{a}
	    \fmfiv{d.sh=circle,d.siz=2thick}{b}
	    \fmfiv{d.sh=circle,d.siz=2thick}{c}
	    \fmfiv{d.sh=circle,d.siz=2thick}{d}
	  \end{fmfgraph}
	}
      \end{fmffile}

      &

      \begin{fmffile}{vp7}
	\fmfframe(1,2)(1,2){ 
	  \begin{fmfgraph}(75,75)
	    \fmfipair{a,b,c,d,e,f}
	    \fmfiequ{a}{.15[nw,ne]}
	    \fmfiequ{b}{.15[sw,se]}
	    \fmfiequ{c}{.85[nw,ne]}
	    \fmfiequ{d}{.85[sw,se]}
	    \fmfiequ{e}{.5[nw,ne]}
	    \fmfiequ{f}{.5[sw,se]}
	    \fmfi{dbl_plain_arrow}{a .. .5[nw,sw] .. b}
	    \fmfi{dbl_plain}{d .. a}
	    \fmfi{dbl_plain_arrow}{d .. .5[e,f]}
	    \fmfi{dbl_plain_arrow}{c .. .5[ne,se] .. d}
	    \fmfi{dbl_plain}{b .. c}
	    \fmfi{dbl_plain_arrow}{b .. .5[e,f]}
	    \fmfi{photon}{a .. .c}
	    \fmfi{dbl_dots}{b .. .d}
	    \fmfiv{d.sh=circle,d.siz=2thick}{a}
	    \fmfiv{d.sh=circle,d.siz=2thick}{b}
	    \fmfiv{d.sh=circle,d.siz=2thick}{c}
	    \fmfiv{d.sh=circle,d.siz=2thick}{d}
	  \end{fmfgraph}
	}
      \end{fmffile}

      &

      \begin{fmffile}{pp6}
	\fmfframe(1,2)(1,2){ 
	  \begin{fmfgraph}(75,75)
	    \fmfipair{a,b,c,d,e,f,g,h,i,j}
	    \fmfiequ{a}{.15[nw,ne]}
	    \fmfiequ{b}{.15[sw,se]}
	    \fmfiequ{c}{.85[nw,ne]}
	    \fmfiequ{d}{.85[sw,se]}
	    \fmfiequ{e}{.3[nw,ne]}
	    \fmfiequ{f}{.3[sw,se]}
	    \fmfiequ{g}{.5[e,f]}
	    \fmfiequ{h}{.7[nw,ne]}
	    \fmfiequ{i}{.7[sw,se]}
	    \fmfiequ{j}{.5[h,i]}
	    \fmfi{dbl_plain_arrow}{a .. .5[nw,sw] .. b}
	    \fmfi{dbl_plain_arrow}{b .. g .. a}
	    \fmfi{dbl_plain_arrow}{c .. .5[ne,se] .. d}
	    \fmfi{dbl_plain_arrow}{d .. j .. c}
	    \fmfi{dbl_dots}{a .. .c}
	    \fmfi{dbl_dots}{b .. .d}
	    \fmfiv{d.sh=circle,d.siz=2thick}{a}
	    \fmfiv{d.sh=circle,d.siz=2thick}{b}
	    \fmfiv{d.sh=circle,d.siz=2thick}{c}
	    \fmfiv{d.sh=circle,d.siz=2thick}{d}
	  \end{fmfgraph}
	}
      \end{fmffile}

      \\
      \\

      &

      \begin{fmffile}{pp7}
	\fmfframe(1,2)(1,2){ 
	  \begin{fmfgraph}(75,75)
	    \fmfipair{a,b,c,d,e,f}
	    \fmfiequ{a}{.15[nw,ne]}
	    \fmfiequ{b}{.15[sw,se]}
	    \fmfiequ{c}{.85[nw,ne]}
	    \fmfiequ{d}{.85[sw,se]}
	    \fmfiequ{e}{.5[nw,ne]}
	    \fmfiequ{f}{.5[sw,se]}
	    \fmfi{dbl_plain_arrow}{a .. .5[nw,sw] .. b}
	    \fmfi{dbl_plain}{d .. a}
	    \fmfi{dbl_plain_arrow}{d .. .5[e,f]}
	    \fmfi{dbl_plain_arrow}{c .. .5[ne,se] .. d}
	    \fmfi{dbl_plain}{b .. c}
	    \fmfi{dbl_plain_arrow}{b .. .5[e,f]}
	    \fmfi{dbl_dots}{a .. .c}
	    \fmfi{dbl_dots}{b .. .d}
	    \fmfiv{d.sh=circle,d.siz=2thick}{a}
	    \fmfiv{d.sh=circle,d.siz=2thick}{b}
	    \fmfiv{d.sh=circle,d.siz=2thick}{c}
	    \fmfiv{d.sh=circle,d.siz=2thick}{d}
	  \end{fmfgraph}
	}
      \end{fmffile}

      &

    \end{tabular}
    \caption{Surviving three-loop diagrams redrawn to show explicitly that they are in fact
    ladder and crossed ladder diagrams (excluding diagrams with the two-pion vertex). 
    Here the double lines represent the baryon and the dashed,
    wavy, and dotted lines correspond to the scalar ($\sigma$), vector ($\omega$), and pion respectively. Notice that 
    there are no scalar-pion and vector-pion ladders because they do not conserve parity.}
    \label{fey10}
  \end{center}
\end{figure}
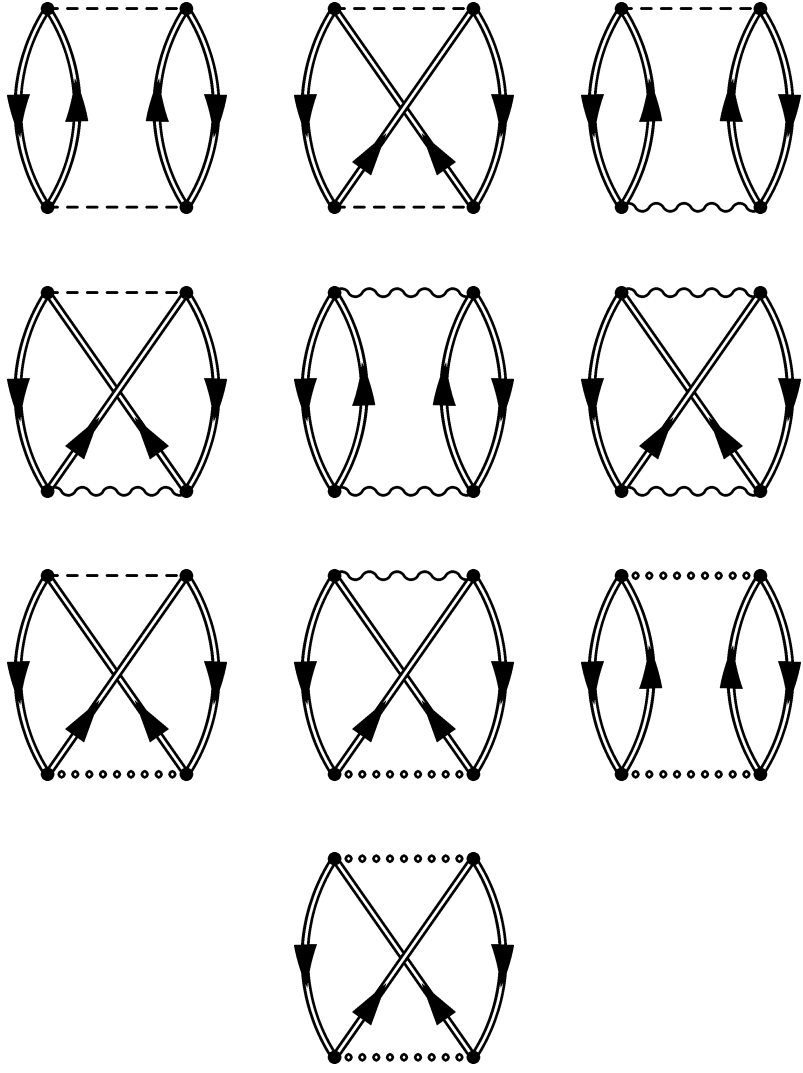

\noindent Note that the propagators used here are defined in \cite{ref:Mc07}.
These integrals correspond to the three diagrams in Fig.\ \ref{fey8}.
The term that produces the third diagram in Fig.\ \ref{fey8} drops out at zero temperature 
because it produces a particle-hole pair of equal momentum at the Fermi surface, which  
violates the Pauli exclusion principle (one can see this more clearly
if the meson propagators are shrunk to contact interactions as in Fig.\ \ref{fey9}) \cite{ref:Fu00a}. 
The first and second diagrams in Fig.\ \ref{fey8} are in fact ladder and crossed ladder diagrams respectively,
as will be seen later. The process by which we have arrived at the above integrals can be repeated for
the various possible combinations of meson propagators. This will yield a ladder and a crossed ladder
for each of following combinations: scalar-scalar, scalar-vector, vector-vector, scalar-pion,
vector-pion, and pion-pion. However, the scalar-pion and vector-pion ladders do not conserve parity
and therefore vanish. This leaves a total of ten integrals (4 ladders and 6 crossed ladders)
which are shown in Fig.\ \ref{fey10}. Note that the above analysis has taken into consideration only the one-pion vertex
[the term proportional to $g_{A} / f_{\pi}$ in Eq.\ (\ref{eqn:lagrangian})]. 

Now, we can write the connected generating functional which contains all the ladders ($L$) and crossed ladders ($XL$)
at the three-loop level. Proceeding as before, the remaining terms are 
(where dimensional regularization was used to make the substitution $G_{H}\rightarrow G^{*}$
\cite{ref:Mc07,ref:Mc07a,ref:Fu89,ref:Se86})
\begin{eqnarray}
{\cal E}_{3}(L+XL) & = & \int\int\frac{d^{4}k}{(2\pi)^{4}}
\frac{d^{4}q}{(2\pi)^{4}}\frac{d^{4}q'}{(2\pi)^{4}}  \nonumber \\
& & \times \left\{-i\frac{g_{S}^{4}}{4}\Delta_{S}(k)\Delta_{S}(k)\tr\left[G^{*}(q-k)G^{*}(q)\right]\right. \nonumber \\[5pt]
& & \quad \times \tr\left[G^{*}(q')G^{*}(q'+k)\right] \nonumber \\[5pt]
& & +i\frac{g_{S}^{4}}{4}\Delta_{S}(k)\Delta_{S}(q'-k)\tr\left[G^{*}(q'+q-k)G^{*}(q) \right. \nonumber \\[5pt]
& & \quad\quad\quad\quad\quad\quad\quad\quad\quad\quad \times \left. G^{*}(q+k)G^{*}(q'+q)\right] \nonumber \\[5pt]
& & +i\frac{g_{S}^{2}g_{V}^{2}}{2}\Delta_{S}(k){\cal D}_{\mu\nu}^{0}(k)
\tr\left[G^{*}(q-k)\gamma_{\mu}G^{*}(q)\right] \nonumber \\[5pt]
& & \quad \times \tr\left[G^{*}(q')\gamma_{\nu}G^{*}(q'+k)\right] \nonumber \\[5pt]
& & -i\frac{g_{S}^{2}g_{V}^{2}}{2}\Delta_{S}(k){\cal D}_{\mu\nu}^{0}(q'-k)
\tr\left[G^{*}(q'+q-k)\gamma_{\mu}G^{*}(q) \right. \nonumber \\[5pt]
& & \quad\quad\quad\quad\quad\quad\quad\quad\quad\quad\quad\; \times \left. G^{*}(q+k)\gamma_{\nu}G^{*}(q'+q)\right] \nonumber \\[5pt]
& & -i\frac{g_{V}^{4}}{4}{\cal D}_{\mu\nu}^{0}(k){\cal D}_{\epsilon\omega}^{0}(k)
\tr\left[\gamma_{\mu}G^{*}(q-k)\gamma_{\omega}G^{*}(q)\right] \nonumber \\[5pt]
& & \quad \times \tr\left[\gamma_{\nu}G^{*}(q')\gamma_{\epsilon}G^{*}(q'+k)\right] \nonumber \\[5pt]
& & +i\frac{g_{V}^{4}}{4}{\cal D}_{\mu\nu}^{0}(k){\cal D}_{\epsilon\omega}^{0}(q'-k)
\tr\left[\gamma_{\mu}G^{*}(q'+q-k)\gamma_{\epsilon}G^{*}(q) \right. \nonumber \\[5pt]
& & \quad\quad\quad\quad\quad\quad\quad\quad\quad\quad\;\;
\times \left. \gamma_{\nu}G^{*}(q+k)\gamma_{\omega}G^{*}(q'+q)\right] \nonumber \\ 
& & +i\frac{g_{S}^{2}}{2}\frac{g_{A}^{2}}{f_{\pi}^{2}}\Delta_{S}(k)\Delta_{\pi}^{ij}(q'-k) \nonumber \\[5pt]
& & \quad \times \tr\left[G^{*}(q'+q-k)(\not\! q'- \not\! k)\frac{\tau_{i}}{2}\gamma_{5}G^{*}(q) \right. \nonumber \\
& & \quad\quad\; \times \left. G^{*}(q+k)(\not\! q' - \not\! k)\frac{\tau_{j}}{2}\gamma_{5}G^{*}(q'+q)\right] \nonumber \\ 
& & -i\frac{g_{V}^{2}}{2}\frac{g_{A}^{2}}{f_{\pi}^{2}}{\cal D}_{\mu\nu}(k)\Delta_{\pi}^{ij}(q'-k) \nonumber \\[5pt]
& & \quad \times \tr\left[\gamma_{\mu}G_{H}(q'+q-k)(\not\! q' - \not\! k)\frac{\tau_{i}}{2}\gamma_{5}G_{H}(q) \right. \nonumber \\
& & \quad\quad\; \times \left. \gamma_{\nu}G_{H}(q+k)(\not\! q' - \not\! k)\frac{\tau_{j}}{2}\gamma_{5}G_{H}(q'+q)\right] \nonumber \\
& & -i\frac{g_{A}^{4}}{4f_{\pi}^{4}}\Delta_{\pi}^{ij}(k)\Delta_{\pi}^{kl}(k)
\tr\left[\not\! k\gamma_{5}\frac{\tau_{i}}{2}G^{*}(q-k)
\not\! k\gamma_{5}\frac{\tau_{k}}{2}G^{*}(q)\right] \nonumber \\
& & \quad \times \tr\left[\not\! k\gamma_{5}\frac{\tau_{j}}{2}G^{*}(q')
\not\! k\gamma_{5}\frac{\tau_{l}}{2}G^{*}(q'+k)\right] \nonumber \\
& & +i\frac{g_{A}^{4}}{4f_{\pi}^{4}}\Delta_{\pi}^{ij}(k)\Delta_{\pi}^{kl}(q'-k) \nonumber \\[5pt]
& & \quad \times \tr\left[\not\! k\gamma_{5}\frac{\tau_{i}}{2}G^{*}(q'+q-k)
(\not\! q' - \not\! k)\gamma_{5}\frac{\tau_{k}}{2}G^{*}(q) \right. \nonumber \\
& & \quad\quad \left.\left. \times \not\! k\gamma_{5}\frac{\tau_{j}}{2}G^{*}(q+k)
(\not\! q' - \not\! k)\gamma_{5}\frac{\tau_{l}}{2}G^{*}(q'+q)\right]\right\} \ . \nonumber \\ & &
\label{eqn:ed3l}
\end{eqnarray}

\noindent In Fig.\ \ref{fey10}, the diagrams corresponding to the integrals in Eq.\ (\ref{eqn:ed3l}) are shown, 
in the same order.

To calculate these ladders and crossed ladders, we must first consider the Bethe-Saltpeter equation.
Two-body correlations can be introduced through an effective interaction $\Gamma$, which is
the solution to the Bethe-Saltpeter equation in the nuclear medium \cite{ref:Br76,ref:Ho87,ref:Sa51}
\begin{equation}
\Gamma=K+\int KGG\Gamma \ ,
\label{eqn:bs}
\end{equation}

\noindent where $K$ is full two-body scattering kernel and $G$ is the fully interacting 
baryon propagator. In practice, the full kernel cannot be written in closed form; we replace 
it with the usual ladder approximation, $K\rightarrow V$. Here $V$ is just a one-meson--exchange 
potential. Then Eq.\ (\ref{eqn:bs}) is rewritten as two equations, or
\begin{eqnarray}
\Gamma & = & U+\int Ug\Gamma \ , \\
U & = & V+\int V\left(GG-g\right)U \ , 
\end{eqnarray}

\noindent where $g$ is an approximate two-body propagator and $U$ is a quasipotential.
For the three-loop case, these equations become
\begin{eqnarray}
\Gamma & = & U+\int UgV \ , \label{eqn:lad1} \\
U & = & V+\int V\left(GG-g\right)U \ . \label{eqn:lad2}
\end{eqnarray}

\noindent If the two-body propagator ($g$) is a good approximation for the product of the 
two interacting baryon propagators, then Eq.\ (\ref{eqn:lad2}) implies $U\approx V$ and Eq.\ (\ref{eqn:lad1})
becomes
\begin{equation}
\Gamma=V+\int VgV \ .
\end{equation}

\noindent Thus we can replace two of the baryon propagators in each of the ladder integrals in 
Eq.\ (\ref{eqn:ed3l}) with a two-body propagator. Specifically, we want to replace the
two propagators that form the interior loop in the ladder diagrams in Fig.\ \ref{fey10};
this turns out to be the loop over the momentum $k$. For the case of the crossed ladders,
we will also insert the two-body propagator into the $k$ loop; here however, we will
also need to perform a Feirz transformation to properly antisymmetrize the integral
\cite{ref:Fi37}.

We replace the two baryon propagators in the $k$ loop with the Blankenbecler-Sugar 
two-particle propagator (assuming $G^{*}G^{*} \sim g$) \cite{ref:Se86,ref:Br76,ref:Ho87,ref:Bl66},
\begin{equation}
g(k,P^{*}|B)=\frac{i}{2}\pi\delta(k_{4})\frac{Q(k,P^{*}|B)\Lambda_{+}^{(1)}
(\frac{1}{2}\vec{P}^{*}+\vec{k})\Lambda_{+}^{(2)}(\frac{1}{2}\vec{P}^{*}-\vec{k})}
{E^{*}(k)\left[{E^{*}}^{2}(k)-\frac{1}{4}s^{*}-ie\right]} \ ,
\label{eqn:tbp1}
\end{equation}

\noindent where the total momentum coming into the $k$ loop is $P_{\mu}$,
$P^{*}_{\mu}=P_{\mu}+2\Sigma_{\mu}$, and $s^{*}={P^{*}}^{2}$ and is assumed to be constant
with respect to the loop momentum \cite{ref:Ho87,ref:Be63}. The $\Sigma_{\mu}$ is
included to eliminate the self energy contributions to the momentum. 
As in the two-loop case, we have
\begin{equation}
E^{*}(k)=\sqrt{\vec{k}^{2}+{M^{*}}^{2}} \ .
\end{equation}

\noindent We also define the following projection operators
\begin{eqnarray}
\Lambda_{+}^{(1)}(\frac{1}{2}\vec{P}^{*}+\vec{k}) 
& = & i\vec{\gamma}\cdot(\frac{1}{2}\vec{P}^{*}+\vec{k})-\gamma_{4}E^{*}(\frac{1}{2}\vec{P}^{*}+\vec{k})-M^{*} \ , \\
\Lambda_{+}^{(2)}(\frac{1}{2}\vec{P}^{*}-\vec{k}) 
& = & i\vec{\gamma}\cdot(\frac{1}{2}\vec{P}^{*}-\vec{k})-\gamma_{4}E^{*}(\frac{1}{2}\vec{P}^{*}-\vec{k})-M^{*} \ .
\end{eqnarray}

\noindent In the nuclear matter limit, we write the Pauli exclusion operators as
\begin{equation}
Q(k,P^{*}|B)=\left[1-\theta\left(k_{F}-|\frac{1}{2}\vec{P}+\vec{k}|\right)\right]
\left[1-\theta\left(k_{F}-|\frac{1}{2}\vec{P}-\vec{k}|\right)\right] \ .
\label{eqn:nn}
\end{equation}

\noindent Here $\vec{P^{*}}=\vec{P}$ (this arises from a relativistic generalization of
the ``reference spectrum'' approximation that is used in nonrelativistic Brueckner calculations \cite{ref:Ho87,ref:Be63}).
Eq.\ (\ref{eqn:nn}) ensures that the two interacting baryon propagators in the $k$ loop are above the Fermi surface.
The propagator in Eq.\ (\ref{eqn:tbp1}) is an approximation for the two baryon propagators that provides the proper
cuts to ensure analyticity and removes the local effects that Infrared Regularization eliminates.
In addition, other versions of the two-body propagator can be used; the alternate
versions considered here are the Thompson two-body propagator \cite{ref:Se86,ref:Th70}
\begin{equation}
g(k,P^{*}|B)=\frac{i}{2}\pi\delta(k_{4})\frac{Q(k,P^{*}|B)\Lambda_{+}^{(1)}
(\frac{1}{2}\vec{P}^{*}+\vec{k})\Lambda_{+}^{(2)}(\frac{1}{2}\vec{P}^{*}-\vec{k})}
{E^{*}(k)\sqrt{-s^{*}}\left[{E^{*}}(k)+\frac{1}{2}\sqrt{-s^{*}}-ie\right]} \ ,
\label{eqn:tbp2}
\end{equation}

\noindent and the Erkelenz-Holinde two-body propagator \cite{ref:Se86,ref:Er72}
\begin{eqnarray}
g(k,P^{*}|B) & = & \frac{i}{2}\pi\delta(k_{4}+\frac{1}{2}\sqrt{-s^{*}}-{E^{*}}(k)) \nonumber \\
& & \times \frac{Q(k,P^{*}|B)\Lambda_{+}^{(1)}
(\frac{1}{2}\vec{P}^{*}+\vec{k})\Lambda_{+}^{(2)}(\frac{1}{2}\vec{P}^{*}-\vec{k})}
{E^{*}(k)\left[{E^{*}}^{2}(k)-\frac{1}{4}s^{*}-ie\right]} \ .
\label{eqn:tbp3}
\end{eqnarray}

Now we are ready to consider the scalar-scalar ladder diagram (the first diagram in Fig.\ \ref{fey10}); 
the box in the center is the loop over $k$. The ladder diagram can be written using the two-body propagator 
in Eq.\ (\ref{eqn:tbp1}) (with $P^{*}_{\mu}=q'_{\mu}+q_{\mu}$)
\begin{eqnarray}
{\cal E}_{3-SS}(L) & = & \frac{g_{S}^{4}\pi}{8}\int\int\frac{d^{4}k}{(2\pi)^{4}}
\frac{d^{4}q}{(2\pi)^{4}}\frac{d^{4}q'}{(2\pi)^{4}}\Delta_{S}(k)\Delta_{S}(k)Q(k,P^{*}|B)  \nonumber \\
& & \times \frac{\delta(k_{4})\tr\left[G^{*}(q)\Lambda_{+}^{(2)}(\frac{1}{2}\vec{P}^{*}-\vec{k})\right]
\tr\left[G^{*}(q')\Lambda_{+}^{(1)}(\frac{1}{2}\vec{P}^{*}+\vec{k})\right]}
{E^{*}(k)\left[{E^{*}}^{2}(k)-\frac{1}{4}s^{*}-ie\right]} \ . \nonumber \\ &&
\end{eqnarray}

\noindent Now we let $G^{*}\rightarrow G^{*}_{F}+G^{*}_{D}$ \cite{ref:Fu89,ref:Se86}. The terms with one or more factor
of $G^{*}_{F}$ will not have enough delta functions to eliminate all the frequency integrals and 
are therefore expressible as a polynomial of terms that are already present in the underlying lagrangian.
These terms are just absorbed and do not need to be calculated.
That leaves us with only the nonlocal portion. We can align the $k$ momentum along the z-axis; 
this allows us to integrate the angular portions of the $k$ integral.
Now, working out the traces (which are over both spin and isospin) and
writing out the angular integrals, we get
\begin{eqnarray}
{\cal E}_{3-SS}(L) & = & -\frac{g_{S}^{4}}{128\pi^{8}}\int_{0}^{k_{F}}\frac{|\vec{q}|^{2}d|\vec{q}|}{E^{*}(q)}
\int_{0}^{k_{F}}\frac{|\vec{q'}|^{2}d|\vec{q'}|}{E^{*}(q')}
\int_{-1}^{1}d(cos\theta_{qk})\int_{-1}^{1}d(cos\theta_{q'k}) \nonumber \\
& & \times \int_{0}^{2\pi}d\phi_{qk}\int_{0}^{2\pi}d\phi_{q'k}
\int \frac{|\vec{k}|^{2}d|\vec{k}|}{E^{*}(k)}Q(k,P^{*}|B) \nonumber \\
& & \times \left[E^{*}(q)E^{*}(\frac{1}{2}\vec{P}^{*}-\vec{k})
+{M^{*}}^{2}-\vec{q}\cdot\left(\frac{1}{2}\vec{P}^{*}-\vec{k}\right)\right] \nonumber \\
& & \times \left[E^{*}(q')E^{*}(\frac{1}{2}\vec{P}^{*}+\vec{k})
+{M^{*}}^{2}-\vec{q'}\cdot\left(\frac{1}{2}\vec{P}^{*}+\vec{k}\right)\right] \nonumber \\
& & / \left\{\left[|\vec{k}|^{2}+m_{S}^{2}\right]^{2}\left[|\vec{k}|^{2}+{M^{*}}^{2}-\frac{1}{4}s^{*}-ie\right]\right\} \ .
\end{eqnarray}

The scalar-scalar crossed ladder diagram (the second diagram in Fig.\ \ref{fey10}) with 
the two-body propagator in Eq.\ (\ref{eqn:tbp1}) is (with $P^{*}_{\mu}=q'_{\mu}+2q_{\mu}$)
\begin{eqnarray}
{\cal E}_{3-SS}(XL) & = & -\frac{g_{S}^{4}\pi}{8}\int\int\frac{d^{4}k}{(2\pi)^{4}}
\frac{d^{4}q}{(2\pi)^{4}}\frac{d^{4}q'}{(2\pi)^{4}}\Delta_{S}(k)\Delta_{S}(q'-k)Q(k,P^{*}|B)  \nonumber \\
& & \times \frac{\delta(k_{4})\tr\left[\Lambda_{+}^{(2)}(\frac{1}{2}\vec{P}^{*}-\vec{k})G^{*}(q)
\Lambda_{+}^{(1)}(\frac{1}{2}\vec{P}^{*}+\vec{k})G^{*}(q'+q)\right]}
{E^{*}(k)\left[{E^{*}}^{2}(k)-\frac{1}{4}s^{*}-i\epsilon\right]} \ . \nonumber \\ && 
\end{eqnarray}

\noindent Now we let $G^{*}\rightarrow G^{*}_{F}+G^{*}_{D}$ \cite{ref:Fu89,ref:Se86}. As before, the terms with one or more factor
of $G^{*}_{F}$ will not have enough delta functions to eliminate all the frequency integrals and 
therefore are expressible as a polynomial of terms that are already present in the underlying lagrangian.
These terms are just absorbed and do not need to be calculated.
That leaves us with only the nonlocal portion. We can align the $k$ momentum along the z-axis; 
this allows us to integrate the angular portions of the $k$ integral. However, the trace is not as 
straightforward as before. In order to use the two-body propagator,
we must also invoke the Fierz identity. This essentially antisymmetrizes the ladder diagram. The Fierz
identity is \cite{ref:Fi37}
\begin{eqnarray}
A \times B & = & \frac{1}{4}\left[\vphantom{\frac{}{}}  AB \times I + A\gamma_{5}B \times \gamma_{5} 
+ A\gamma_{\alpha}B \times \gamma_{\alpha}\right. \nonumber \\
& & \left. - A\gamma_{\alpha}\gamma_{5}B \times \gamma_{\alpha}\gamma_{5} 
+ \frac{1}{2}A\sigma_{\alpha\beta}B \times \sigma_{\alpha\beta}\right] \ ,
\end{eqnarray}

\noindent where $A$ and $B$ represent either $\gamma_{\mu}$ or $I$ and the $\times$ represents any matrices
that come between $A$ and $B$. Substituting this into the scalar-scalar crossed ladder integral, we get
\begin{eqnarray}
{\cal E}_{3-SS}(XL) & = & \frac{g_{S}^{4}}{2^{8}\pi^{8}}\int_{0}^{k_{F}}\frac{|\vec{q}|^{2}d|\vec{q}|}{E^{*}(q)}
\int_{-1}^{1}d(cos\theta_{qk})\int_{-1}^{1}d(cos\theta_{q'k})\int_{0}^{2\pi}d\phi_{qk}\int_{0}^{2\pi}d\phi_{q'k}  \nonumber \\
& & \times \int\frac{|\vec{q'}|^{2}d|\vec{q'}|}{E^{*}(q'+q)}\theta\left(k_{F}-|\vec{q'}+\vec{q}|\right)
\int \frac{|\vec{k}|^{2}d|\vec{k}|}{E^{*}(k)}Q(k,P^{*}|B) \nonumber \\
& & \times \left[\left(\frac{1}{2}\vec{P}^{*}-\vec{k}\right)\cdot\left(\vec{q'}+\vec{q}\right)
-E^{*}(\frac{1}{2}\vec{P}^{*}-\vec{k})E^{*}(q'+q)-{M^{*}}^{2}\right] \nonumber \\
& & \times \left[\left(\frac{1}{2}\vec{P}^{*}+\vec{k}\right)\cdot \vec{q}
-E^{*}(\frac{1}{2}\vec{P}^{*}+\vec{k})E^{*}(q)-{M^{*}}^{2}\right] \nonumber \\
& & / \left\{\left[|\vec{k}|^{2}+m_{S}^{2}\right]
\left[\left(|\vec{q'}|-|\vec{k}|\right)^{2}-\left[E^{*}(q'+q)-E^{*}(q)\right]^{2}+m_{S}^{2}\right] \right. \nonumber \\
& & \times \left.\left[|\vec{k}|^{2}+{M^{*}}^{2}-\frac{1}{4}s^{*}-ie\right]\right\} \ . 
\end{eqnarray}

\noindent The other ladders and crossed ladders are handled in a similar fashion.
The only difference is the traces are more complicated and the coupling constants change.
In addition, one can substitute Eqs.\ (\ref{eqn:tbp2}) and (\ref{eqn:tbp3}) for Eq.\ (\ref{eqn:tbp1}).

\subsection{Two-pion Vertex}

In this subsection, we consider the contributions to the loop expansion at the three-loop level
involving the two-pion vertex. These terms arise from the second pion-nucleon coupling in 
Eq.\ (\ref{eqn:lagrangian}). There are two possible combinations that conserve isopsin
and parity (and are not redundant); this corresponds to the following portion of the connected generating functional:
\begin{eqnarray}
\delta W_{3-\pi\pi} & = & \left\{-i\frac{\hbar^{3}}{2}\frac{g_{A}^{2}}{2f_{\pi}^{4}} 
\int\int d^{4}xd^{4}yd^{4}zd^{4}z' \right. \nonumber \\
& & \quad \times \left(\gamma_{\mu}\gamma_{5}\partial_{\mu}^{x}\left[\frac{-i\delta}{\delta \zeta_{i}(x)}\right]
\cdot\frac{\tau_{i}}{2}\right)_{\alpha\alpha'}
\left(\gamma_{\nu}\gamma_{5}\partial_{\nu}^{y}\left[\frac{-i\delta}{\delta \zeta_{j}(y)}\right]
\cdot\frac{\tau_{j}}{2}\right)_{\beta\beta'}  \nonumber \\
& & \quad \times \left\{\gamma_{\epsilon}\left[\frac{-i\delta}{\delta \zeta_{k}(z)}\right]
\left(\partial_{\epsilon}^{z'}\left[\frac{-i\delta}{\delta \zeta_{l}(z')}\right]\right)
\left[\frac{\tau_{k}}{2},\frac{\tau_{l}}{2}\right]\delta_{zz'}\right\}_{\gamma\gamma'} \nonumber \\
& & \quad \times \left[\frac{i\delta}{\delta \xi(x)}\right]_{\alpha}\left[\frac{-i\delta}{\delta \bar{\xi}(x)}\right]_{\alpha'}
\left[\frac{i\delta}{\delta \xi(y)}\right]_{\beta}\left[\frac{-i\delta}{\delta \bar{\xi}(y)}\right]_{\beta'} 
\left[\frac{i\delta}{\delta \xi(z)}\right]_{\gamma}\left[\frac{-i\delta}{\delta \bar{\xi}(z)}\right]_{\gamma'} \nonumber \\ 
& & -i\frac{\hbar^{3}}{2}\frac{1}{4f_{\pi}^{4}} \int\int d^{4}xd^{4}x'd^{4}yd^{4}y' \nonumber \\
& & \quad \times \left\{\gamma_{\mu}\left[\frac{-i\delta}{\delta \zeta_{i}(x)}\right]
\left(\partial_{\mu}^{x'}\left[\frac{-i\delta}{\delta \zeta_{j}(x')}\right]\right)
\left[\frac{\tau_{i}}{2},\frac{\tau_{j}}{2}\right]\delta_{xx'}\right\}_{\alpha\alpha'} \nonumber \\
& & \quad \times \left\{\gamma_{\nu}\left[\frac{-i\delta}{\delta \zeta_{k}(y)}\right]
\left(\partial_{\nu}^{y'}\left[\frac{-i\delta}{\delta \zeta_{l}(y')}\right]\right)
\left[\frac{\tau_{k}}{2},\frac{\tau_{l}}{2}\right]\delta_{yy'}\right\}_{\beta\beta'} \nonumber \\
& & \quad \left. \times \left[\frac{i\delta}{\delta \xi(x)}\right]_{\alpha}\left[\frac{-i\delta}{\delta \bar{\xi}(x)}\right]_{\alpha'}
\left[\frac{i\delta}{\delta \xi(y)}\right]_{\beta}\left[\frac{-i\delta}{\delta \bar{\xi}(y)}\right]_{\beta'}\right\} \nonumber \\ 
& & \times \exp\left\{-i \int\int d^{4}x_{1}d^{4}x_{2} \bar{\xi}(x_{1})G_{H}(x_{1}-x_{2})\xi(x_{2})\right\} \nonumber \\ 
& & \left. \times \exp\left\{\frac{i}{2}\int\int d^{4}x_{1}d^{4}x_{2}\zeta_{c}(x_{1})\Delta_{\pi}^{cd}(x_{1}-x_{2})\zeta_{d}(x_{2})\right\}
\right|_{sources=0} \nonumber \\ & & - VEV \ .
\end{eqnarray}

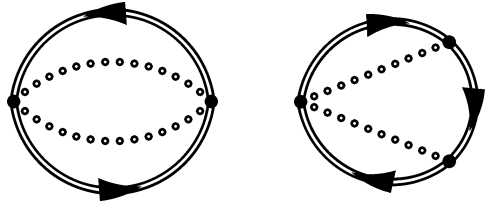
\begin{figure}[!htb]
  \begin{center}
    \begin{tabular}{cc}
      \begin{fmffile}{epp1}
	\fmfframe(1,2)(1,2){ 
	  \begin{fmfgraph}(75,75)
	    \fmfipair{a,b,c,d,e,f,g,h}
	    \fmfiequ{a}{.5[nw,ne]}
	    \fmfiequ{b}{.5[nw,sw]}
	    \fmfiequ{c}{.5[sw,se]}
	    \fmfiequ{d}{.5[se,ne]}
	    \fmfiequ{e}{.7[a,c]}
	    \fmfiequ{f}{.3[a,c]}
	    \fmfiequ{g}{.95[a,c]}
	    \fmfiequ{h}{.05[a,c]}
	    \fmfi{dbl_plain_arrow}{b .. g  .. d}
	    \fmfi{dbl_plain_arrow}{d .. h  .. b}
	    \fmfi{dbl_dots}{b .. e .. d}
	    \fmfi{dbl_dots}{b .. f .. d}
	    \fmfiv{d.sh=circle,d.siz=2thick}{b}
	    \fmfiv{d.sh=circle,d.siz=2thick}{d}
	  \end{fmfgraph}
	}
      \end{fmffile}

      &

      \begin{fmffile}{epp2}
	\fmfframe(1,2)(1,2){ 
	  \begin{fmfgraph}(75,75)
	    \fmfipair{a,b,c,d,e,f,g,h,i,j,k,l,m,n}
	    \fmfiequ{a}{.5[nw,ne]}
	    \fmfiequ{b}{.5[nw,sw]}
	    \fmfiequ{c}{.5[sw,se]}
	    \fmfiequ{d}{.5[se,ne]}
	    \fmfiequ{e}{.75[nw,ne]}
	    \fmfiequ{f}{.75[sw,se]}
	    \fmfiequ{g}{.75[b,d]}
	    \fmfiequ{h}{.4[e,g]}
	    \fmfiequ{i}{.4[f,g]}
	    \fmfiequ{j}{.88[b,d]}
	    \fmfiequ{k}{.4[nw,ne]}
	    \fmfiequ{l}{.4[sw,se]}
	    \fmfiequ{m}{.9[k,l]}
	    \fmfiequ{n}{.1[k,l]}
	    \fmfi{dbl_plain_arrow}{b .. n .. h}
	    \fmfi{dbl_plain_arrow}{h .. j .. i}
	    \fmfi{dbl_plain_arrow}{i .. m .. b}
	    \fmfi{dbl_dots}{b .. h}
	    \fmfi{dbl_dots}{b .. i}
	    \fmfiv{d.sh=circle,d.siz=2thick}{b}
	    \fmfiv{d.sh=circle,d.siz=2thick}{h}
	    \fmfiv{d.sh=circle,d.siz=2thick}{i}
	  \end{fmfgraph}
	}
      \end{fmffile}
    \end{tabular}
    \caption{Three-loops diagrams involving the two-pion vertex.}
    \label{fey11}
  \end{center}
\end{figure}

\noindent These terms are represented by the diagrams in Fig.\ \ref{fey11}. After working out the variational derivatives,
taking the Fourier transforms, and using all the delta functions, the term corresponding to the second
diagram in Fig.\ \ref{fey11} drops out. As the two-pion vertex is antisymmetric, only a combination
that includes two of these vertices can survive. The remaining (football) diagram 
is represented by the following integral (after transforming to the energy density and suppressing
the $\hbar$)
\begin{eqnarray}
{\cal E}_{3-\pi\pi}(F) & = & -i\frac{9}{8f_{\pi}^{4}} \int\int \frac{d^{4}k}{(2\pi)^{4}}
\frac{d^{4}q}{(2\pi)^{4}}\frac{d^{4}q'}{(2\pi)^{4}}\Delta_{\pi}(k)\Delta_{\pi}(q') \nonumber \\
& & \times \Tr\left\{\left(\not\! q'-\not\! k\right)G^{*}(q)
\not\! k\; G^{*}(q-k-q')\right\} \ ,
\end{eqnarray}

\noindent where Tr is summed only over the spin (the isospin sum has already been done).
Again, dimensional regularization was used to make the substitution $G_{H}\rightarrow G^{*}$
\cite{ref:Mc07,ref:Mc07a,ref:Fu89,ref:Se86}.

We now proceed in a manner similar to the method used in section \ref{sec:1a}.
Here we need to introduce a two-pion propagator ($\Delta_{\pi}\Delta_{\pi} \sim \Delta_{\pi\pi}$),
\begin{equation}
\Delta_{\pi\pi}(k,P^{*})=\frac{i}{2}\pi\delta(k_{4})\frac{1}
{E^{*}_{\pi}(k)\left[{E^{*}_{\pi}}^{2}(k)-\frac{1}{4}s^{*}-ie\right]} \ ,
\end{equation}

\noindent where 
\begin{equation}
E^{*}_{\pi}(k)=\sqrt{|\vec{k}|^{2}+m^{2}_{\pi}} \ .
\end{equation}

\noindent The two-pion propagator follows from the two-body Blankenbecler-Sugar nucleon
propagator without the projection operators and the Heavyside step functions \cite{ref:Br76,ref:Bl66}.
Thompson and Erkelenz-Holinde forms can also be used.
This allows us to rewrite the integral as (with $P_{\mu}=q'_{\mu}$)
\begin{eqnarray}
{\cal E}_{3-\pi\pi}(F) & = & -\frac{9\pi^{3}}{16f_{\pi}^{4}} \int\int \frac{d^{4}k}{(2\pi)^{4}}
\frac{d^{4}q}{(2\pi)^{4}}\frac{d^{4}q'}{(2\pi)^{4}}\delta(k_{4}) \nonumber \\[5pt]
& & \times \Tr\left[\left(\not\! q'-2\not\! k\right)G^{*}(q)\not\! k\; G^{*}(q'+q)\right] \nonumber \\
& & / \left\{E^{*}_{\pi}(k)\left[{E^{*}_{\pi}}^{2}(k)-\frac{1}{4}s^{*}-ie\right]\right\} \ .
\end{eqnarray}

\noindent Now we let $G^{*}\rightarrow G^{*}_{F}+G^{*}_{D}$ \cite{ref:Fu89,ref:Se86}. 
As before, the terms with one or more factor
of $G^{*}_{F}$ will not have enough delta functions to eliminate all the frequency integrals and 
therefore are expressible as a polynomial of terms that are already present in the underlying lagrangian.
These terms are just absorbed and do not need to be calculated.
That leaves us with only the nonlocal portion. We can align the $k$ momentum along the z-axis; 
this allows us to integrate the angular portions of the $k$ integral.
Now, working out the trace (which is only over spin) and
writing out the angular integrals, we get
\begin{eqnarray}
& = & \frac{9}{2^{12}\pi^{8}f_{\pi}^{4}} \int_{0}^{k_{F}}\frac{|\vec{q}|^{2}d|\vec{q}|}{E^{*}(q)}
\int_{-1}^{1}d(cos\theta_{qk})\int_{-1}^{1}d(cos\theta_{q'k})\int_{0}^{2\pi}d\phi_{qk}\int_{0}^{2\pi}d\phi_{q'k}  \nonumber \\
& & \times \int\frac{|\vec{q'}|^{2}d|\vec{q'}|}{E^{*}(q'+q)}\theta\left(k_{F}-|\vec{q'}+\vec{q}|\right)
\int \frac{|\vec{k}|^{2}d|\vec{k}|}{E^{*}_{\pi}(k)} \nonumber \\
& & \times \left(\vec{k}^{2}\left\{2E^{*}(q)\left[E^{*}(q'+q)-E^{*}(q)\right]
-2\vec{q'}\cdot\vec{q}\right\} \right. \nonumber \\
& & \quad \left. +\vec{q}\cdot\vec{k}\left\{4\vec{q}\cdot\vec{k}-2\vec{q'}\cdot\vec{q}+4\vec{q'}\cdot\vec{k}
-\vec{q'}^{2} \right.\right. \nonumber \\
& & \quad\quad\quad \left.\left. +\left[E^{*}(q'+q)-E^{*}(q)\right]\left[E^{*}(q'+q)+E^{*}(q)\right]
\vphantom{\vec{q'}} \right\}\right) \nonumber \\
& & / \left\{{E^{*}_{\pi}}^{2}(k)-\frac{1}{4}\vec{q'}^{2}+\frac{1}{4}\left[E^{*}(q'+q)-E^{*}(q)\right]^{2}\right\} \ .
\end{eqnarray}

\section{Discussion}
\label{sec:2}

In this section, we discuss the results of the three-loop calculations for QHD. The level of
truncation retained is sufficient to study the general effects of the three-loop energy (as can be seen at the two-loop
order in \cite{ref:Mc07,ref:Mc07a}). 
No nonlinear meson self-interactions were included in the underlying lagrangian.
We leave the inclusion of these nonlinear effects for 
future work. In addition, it was shown implicitly that the local, short-range dynamics
were absorbed into the parameterization of the lagrangian (as it is expressible as
a series of terms that in principle already exist). What remains are the nonlocal,
long-range correlations, which are calculated explicitly.

\begin{figure}
\begin{center}
\includegraphics[width=4 in]{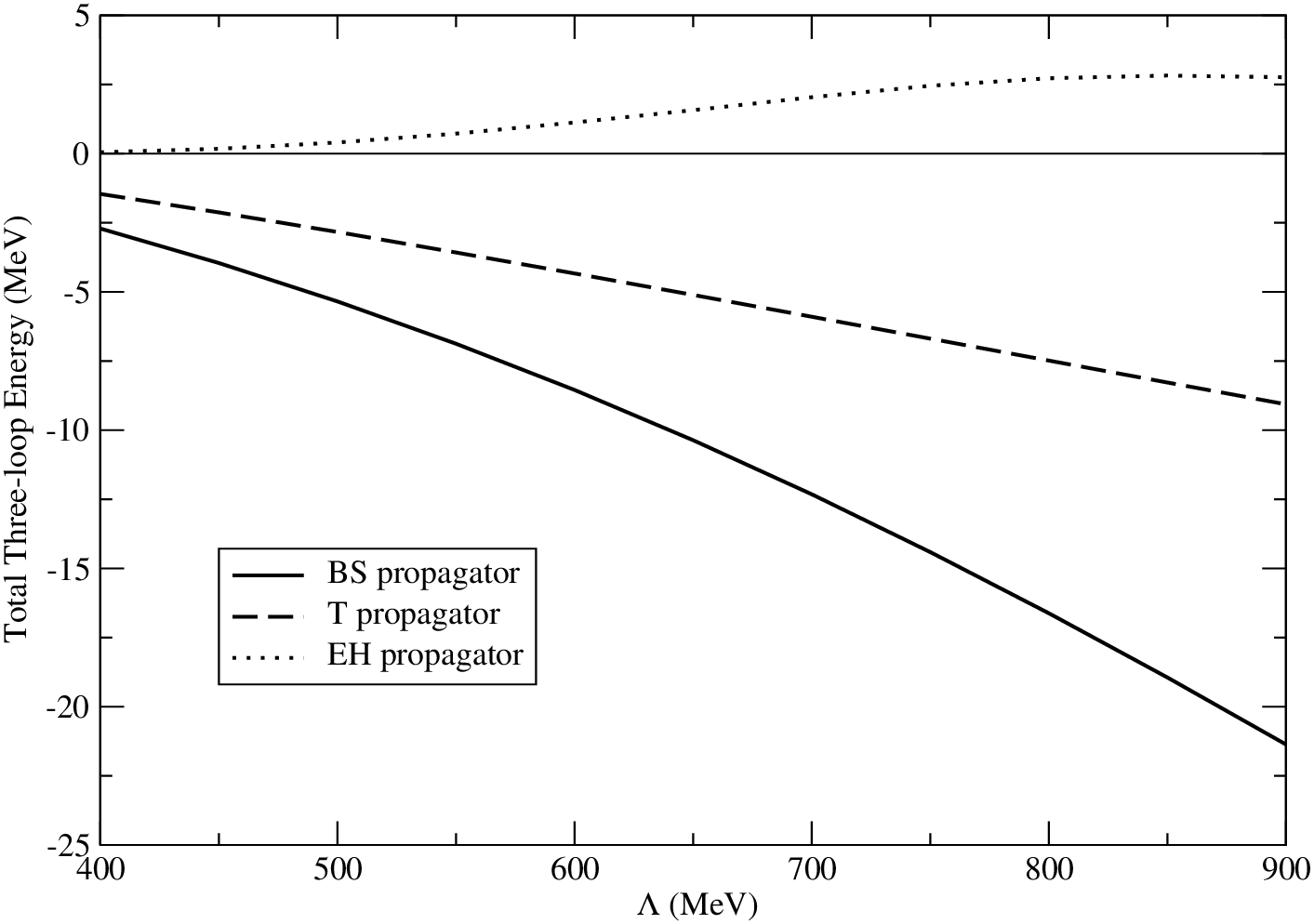}
\caption{The total three-loop energy density as a function of cutoff. Three versions
of the two-body propagator were investigated: Blankenbecler-Sugar (BS), Thompson (T), and
Erkelenz-Holinde (EH). Here the M0A parameter set was used \cite{ref:Mc07}.}
\label{fig:3loop1}
\end{center}
\end{figure}

Now we consider the numerical analysis of the 
three-loop energy. We use the parameter sets
listed in Table \ref{tab:2}. For the pion terms, we use
$g_{A}^{2}=1.5876$ and $f_{\pi}=93$ MeV \cite{ref:Fu97}. In
addition, both the W1 (one-loop level) \cite{ref:Fu97} and M0A (two-loop level) 
\cite{ref:Mc07} sets lead to equilibrium at $k_{\mathrm
F}=1.3$ fm$^{-1}$. The mean meson fields are determined by
extremizing the meson field equations and are used as input to
the exchange integrals. Then the full three-loop energy density is
extremized with respect to the meson fields.
The two-loop integrals were calculated using Gaussian quadrature.

A total of eleven integrals survive at the three-loop level: four ladders, six crossed ladders, and one 
football diagram. The three-loop integrals are also solved using Gaussian quadrature.
However, the interior loop (the one over momentum $k$) in all eleven integrals
has an upper limit of infinity. As a result, a cutoff must be introduced. 
This is expected, since the effective field theory is applicable only up to 
some breakdown scale determined by non-Goldstone boson physics. The 
sum of the three-loop contributions to the energy density as a function of cutoff is
shown in Fig.\ \ref{fig:3loop1}. The results in Fig.\ \ref{fig:3loop1} are for three different versions of the two-body 
propagator [see Eqs.\ (\ref{eqn:tbp1}), (\ref{eqn:tbp2}), and (\ref{eqn:tbp3})] 
and utilize the M0A parameter set \cite{ref:Mc07}. 
The choice of the cutoff is critical.
If the cutoff is too high, then some local physics gets included
and the energy becomes large. If the cutoff is too small, some of the
nonlocal dynamics is lost. A reasonable cutoff 
($600 \lsim \Lambda \lsim 700$ MeV \cite{ref:Fu04,ref:Vr04}) yields a total three-loop energy of less than $17$ MeV. For 
comparison, the total two-loop
energy for the M0A set is $29.60$ MeV. One can see that for a reasonable choice of cutoff,
the energy decreases from the two- to three-loop order. 

It is clear from Fig.\ \ref{fig:3loop1} that there is some uncertainty arising from the two-body
propagators. To get an improved idea of how good an 
approximation each propagator is, Eqs.\ (\ref{eqn:lad1}) and (\ref{eqn:lad2})
need to be solved self-consistently. This is left for future work.

In Table \ref{tab:1}, the individual
contributions are shown using the M0A set for four choices of the cutoff 
($\Lambda = 500$, $600$, $700$, and $800$ MeV) and the Blankenbecler-Sugar propagator.
The terms are quite large; however, there
is always some cancellation between ladders and their corresponding crossed ladders. In addition,
there is large cancellation when the scalar-scalar, scalar-vector, and vector-vector terms
are summed. 

\begin{figure}
\begin{center}
\includegraphics[width=4 in]{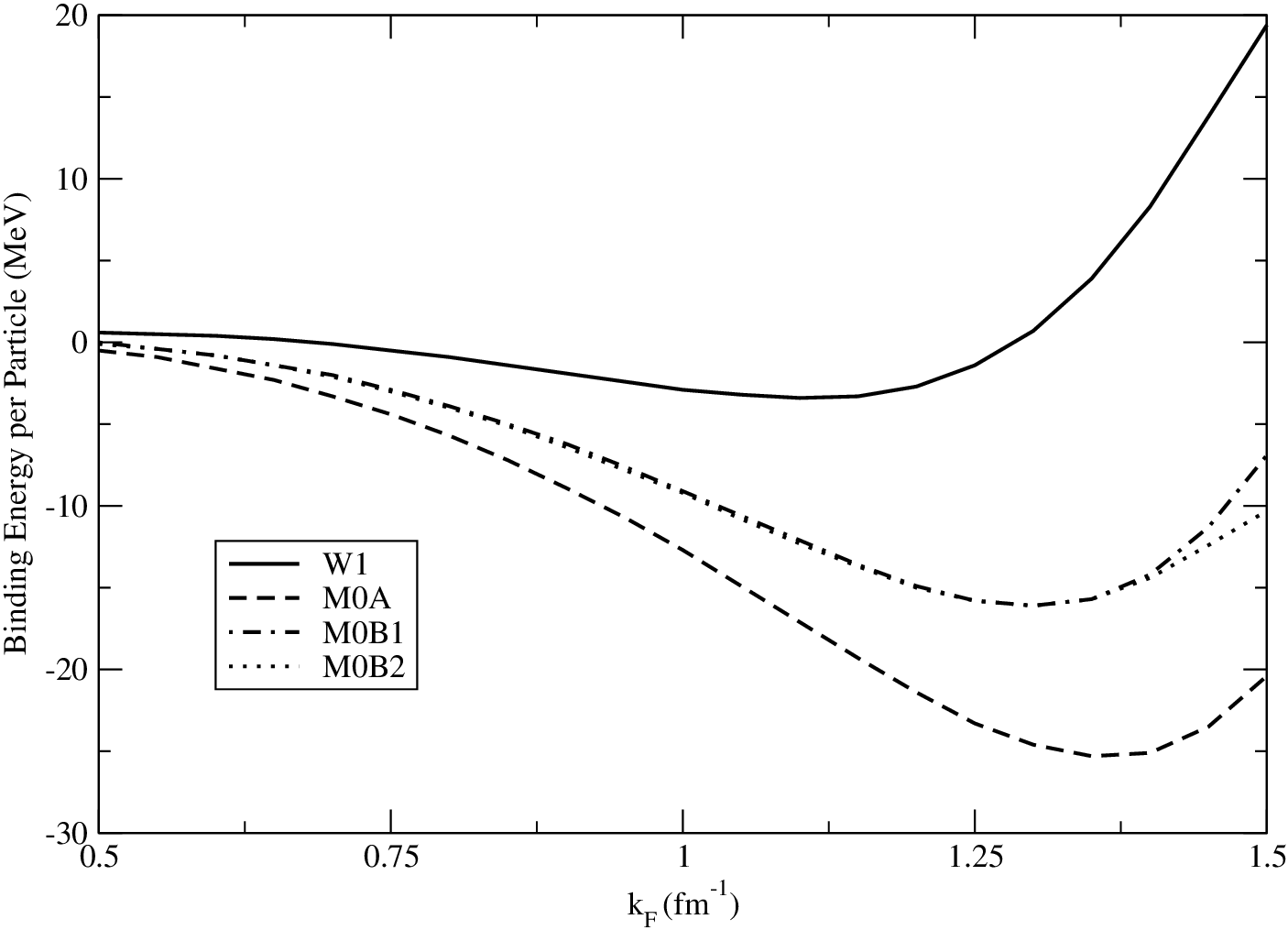}
\caption{Comparison of the nuclear binding curves for the set W1 (one-loop parameters but 
with the two- and three-loop contributions included and $\Lambda=600$), M0A (two-loop parameters but 
with the three-loop contributions included and $\Lambda=600$), M0B1 (three-loop parameters with $\Lambda=600$),
and M0B2 (three-loop parameters with $\Lambda=700$).\vspace{.1 in}}
\label{fig:3loop4}
\end{center}
\end{figure}

The set M0A listed in Table \ref{tab:2} was fit at the two-loop
level to nuclear equilibrium (${\cal E}/\rho_{B}-M=-16.1$ MeV and
$k_{\mathrm F}=1.30$ fm$^{-1}$) by adjusting $g_S$ and $g_V$ using a
downhill simplex method to minimize a least-squares fit (with
respective weights of $0.0015$ and $0.002$) \cite{ref:Mc07}. The sets
M0B1 and M0B2 (also in Table \ref{tab:2}) were fit at the three-loop level to nuclear equilibrium using the same method 
and the Blankenbecler-Sugar propagator for cutoffs of 600 and 700 MeV
respectively. The binding curves for W1, M0A, M0B1, and M0B2 are
shown in Fig.\ \ref{fig:3loop4}. All four curves are for the total energy up to the three-loop
level and have been extremized in the both heavy meson fields.
One can see that while the two- and three-loop level contributions
are not large, they are not negligible either. Furthermore, 
nuclear saturation can be reproduced at the three-loop level for a range
of cutoffs with parameter sets that are natural ($g_{S,V} \approx 4\pi$). 
Note that the natural size of the fitted $g_{S}$ implies that scalar-exchange 
diagrams are {\it not} redundant. 

\begin{table}
\begin{center}
\begin{tabular}{|c|c|c|c|c|} \hline
             & W1 \cite{ref:Fu97} & M0A \cite{ref:Mc07} & M0B1        & M0B2        \\ \hline 
$\Lambda$    & ---                & ---                 & 600         & 700         \\ \hline
$m_{S}/M$    & 0.60305            & $0.5400\xx$         & $0.5400\xx$ & $0.5400\xx$ \\ \hline 
$m_{V}/M$    & 0.83280            & 0.83280             & 0.83280     & 0.83280     \\ \hline 
$g_{S}/4\pi$ & 0.93797            & 0.79361             & 0.79970     & 0.81819     \\ \hline 
$g_{V}/4\pi$ & 1.13652            & 0.96811             & 1.01496     & 1.06487     \\ \hline
\end{tabular}
\caption{Parameter sets used in this work. Here $\Lambda$ is in MeV.} 
\label{tab:2}
\end{center}
\end{table}

\begin{table}
\begin{center}
\begin{tabular}{|l|c|c|c|c|} \hline
$\xx\xx\xx\xx\xx\Lambda$      & $500$                 & $600$                 & $700$                 
& $800$                 \\ \hline\hline 
${\cal E}_{3-SS}(L)$          & $-20.80$              & $-28.54$              & $-35.15$              
& $-40.62$              \\ \hline 
${\cal E}_{3-SS}(XL)$         & $\zz\xx\xx 5.038$     & $\zz\xx\xx 7.658$     & $\zz\xx\xx 9.921$     
& $\zz 11.81$           \\ \hline 
${\cal E}_{3-SV}(L)$          & $\zz 36.10$           & $\zz 52.73$           & $68.45$               
& $\zz 82.69$           \\ \hline 
${\cal E}_{3-SV}(XL)$         & $\xx -9.389$          & $-14.94$              & $-20.22$              
& $-25.04$              \\ \hline
${\cal E}_{3-VV}(L)$          & $-16.97$              & $-27.08$              & $-38.02$              
& $-49.23$              \\ \hline 
${\cal E}_{3-VV}(XL)$         & $\zz\xx\xx 4.438$     & $\zz\xx\xx 7.674$     & $\zz 11.19$           
& $\zz 14.80$           \\ \hline 
${\cal E}_{3-S\pi}(XL)$       & $\zz\xx\xx\xx 0.6105$ & $\zz\xx\xx 1.193$     & $\zz\xx\xx 1.998$     
& $\zz\xx\xx 2.997$     \\ \hline
${\cal E}_{3-V\pi}(XL)$       & $\xx -1.127$          & $\xx -1.544$          & $\xx -1.598$          
& $\xx -1.216$          \\ \hline 
${\cal E}_{3-\pi\pi}(L)$      & $\xx -5.213$          & $\xx -9.503$          & $-15.05$              
& $-21.84$              \\ \hline 
${\cal E}_{3-\pi\pi}(XL)$     & $\zz\xx\xx 1.705$     & $\zz\xx\xx 3.411$     & $\zz\xx\xx 5.617$     
& $\zz\xx\xx 8.292$     \\ \hline
${\cal E}_{3-\pi\pi}(F)$      & $\zz\xx\xx\xx 0.2618$ & $\zz\xx\xx\xx 0.3933$ & $\zz\xx\xx\xx 0.5507$ 
& $\zz\xx\xx\xx 0.7361$ \\ \hline\hline
$\xx\xx\xx\xx\xx{\cal E}_{3}$ & $\xx -5.345$          & $\xx -8.548$          & $-12.32$              
& $-16.62$              \\ \hline
\end{tabular}
\caption{Individual contributions to the three-loop energy density from ladders ($L$),
crossed ladders ($XL$), and football ($F$) terms for a range of cutoffs
using the Blankenbecler-Sugar propagator and the M0A set at nuclear equilibrium \cite{ref:Mc07}.
All entries are in MeV.} 
\label{tab:1}
\end{center}
\end{table}

\begin{figure}
\begin{center}
\includegraphics[width=4 in]{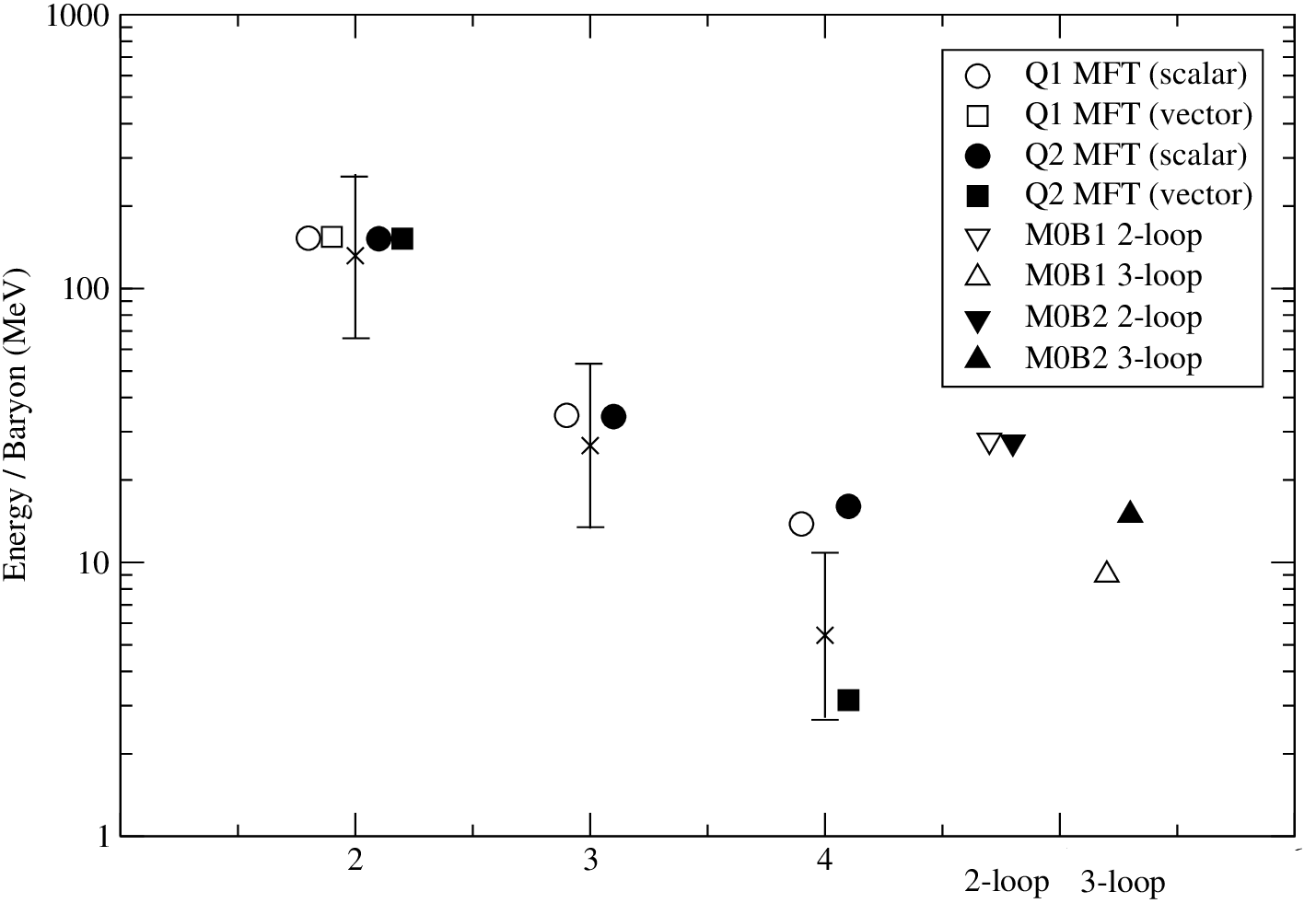}
\caption{Comparison of the magnitudes of the mean field terms in the
meson sector with the two- and three-loop exchange integrals. The inverted
triangle represents the total two-loop contribution and the
triangle represents the total three-loop contribution. The abscissa
represents the order $\nu$ in the power counting at the mean field
level \cite{ref:Fu97}. Absolute values are shown.\vspace{.1 in}}
\label{fig:3loop5}
\end{center}
\end{figure}

Now we want to see how the three-loop energy fits in phenomenologically with the power
counting. Table \ref{tab:1} shows clearly that the individual terms may be large, 
but strong cancellation occurs. Thus, the proper comparison is with the total energy at each order in the loops.
Fig.\ \ref{fig:3loop5} plots the magnitude of the energy for both the mean field power counting
hierarchy \cite{ref:Fu97,ref:Fu03} along side the two- and three-loop energies. Here the three-loop
energies utilize the Blankenbecler-Sugar propagator and cutoffs of 600 and 700 MeV. 
The total two-loop energy is equivalent to third order in the mean
field power counting, while the three-loop energy is comparable to fourth order. The same
hierarchy is observed for the Thompson and Erkelenz-Holinde propagators (in fact it is more pronounced,
as is clear from Fig.\ \ref{fig:3loop1}). Here the mean field
parameter sets used (Q1 and Q2 \cite{ref:Fu97}) include various nonlinearities in the
isoscalar meson fields; we stress that these nonlinearities were not included in the present
analysis of the two- and three-loop energies and were shown just to illustrate the scale of the
mean-field nonlinear terms.

\begin{figure}
\begin{center}
\includegraphics[width=4 in]{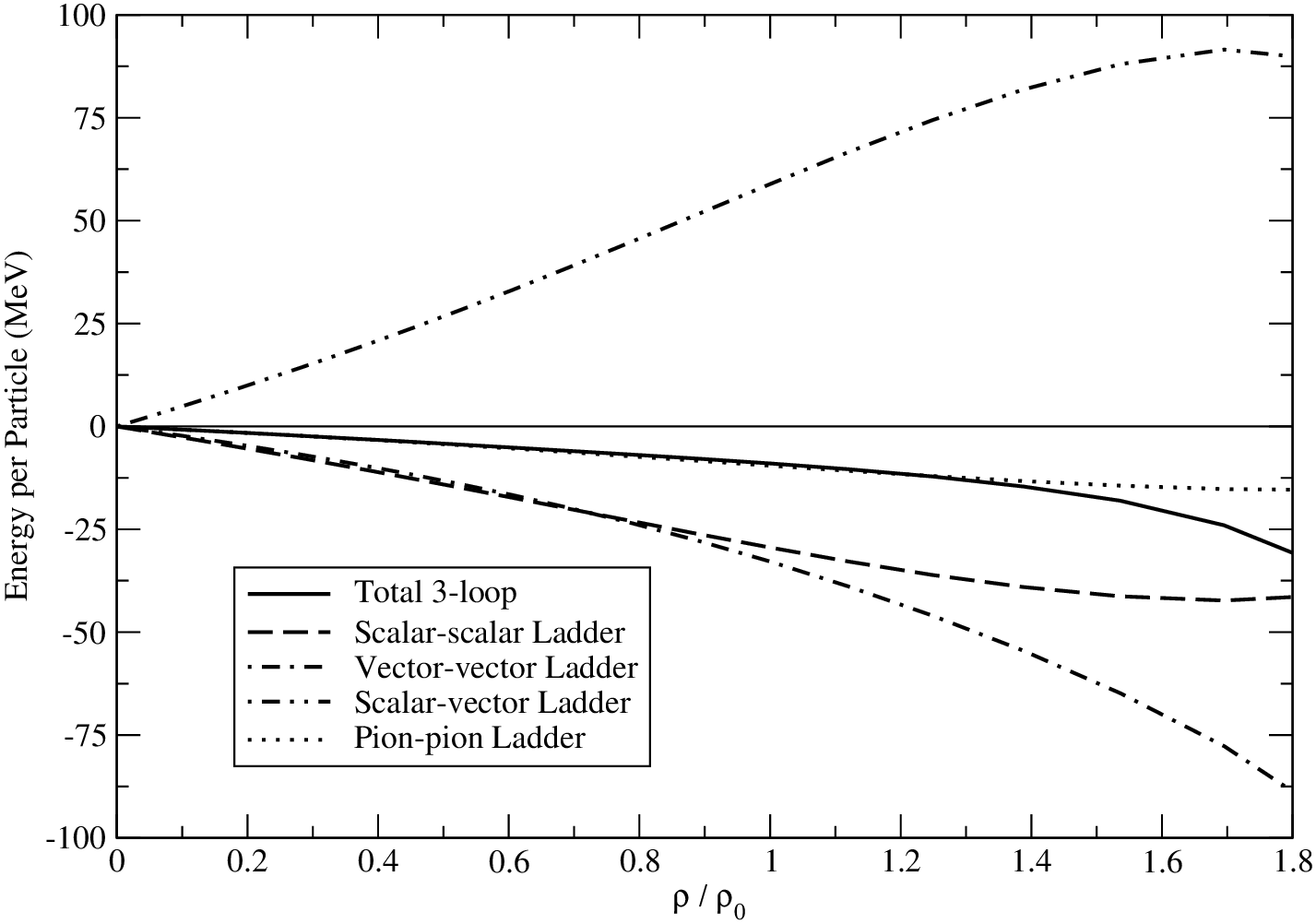}
\caption{Behavior of ladder integrals as a function of density using the Blankenbecler-Sugar
propagator. Here the M0B1 parameter set was used.\vspace{.2 in}}
\label{fig:3loop2}
\end{center}
\end{figure}

\begin{figure}
\begin{center}
\includegraphics[width=4 in]{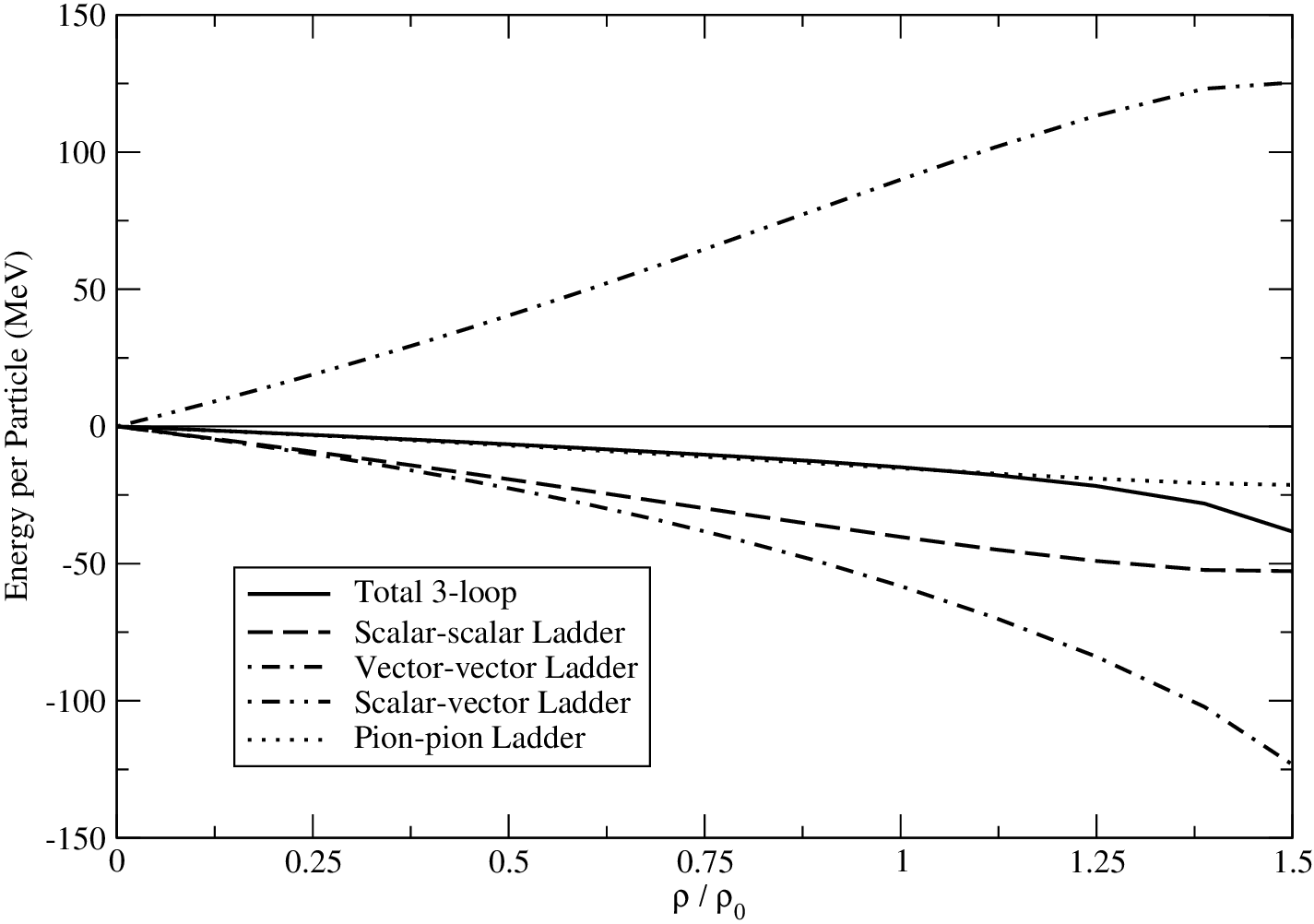}
\caption{Behavior of ladder integrals as a function of density using the Blankenbecler-Sugar
propagator. Here the M0B2 parameter set was used.}
\label{fig:3loop3}
\end{center}
\end{figure}

It is also of interest to investigate how the individual and total three-loop contributions behave with respect 
to density. Figs.\ \ref{fig:3loop2} and \ref{fig:3loop3} graph the individual ladder integrals and
the total three-loop energy as a function of density ($\rho/\rho_{0}$ where $\rho = \rho_{B}$)
for the cutoffs 600 and 700 MeV respectively. One can see that while the total three-loop energy is roughly
linear, the individual ladder contributions are nonlinear.

It is worthy of note that these results disagree with the claim in \cite{ref:Vr04}, that the two- and
three-loop pion-exchange terms are critical for providing nuclear binding and saturation. This
follows naturally, since the two-loop pion exchange terms are repulsive, and the (attractive)
three-loop terms are smaller than the two-loop terms for reasonable values of the cutoff $\Lambda$.
The sum of the pure pion terms for the M0B1 and M0B2 sets is $+6.7$ and $+3.3$ MeV respectively.


In {\it summary}, we have conducted the loop expansion to third-order for QHD.
Since covariant many-body loop expansions have been studied for more than thirty years \cite{ref:Ch77},
it is important to enumerate the new features of this work. First, the loop expansion was
conducted while incorporating {\it both} the chiral and heavy meson dynamics. Second, it was shown that,
for a reasonable choice of cutoff, the energy contributions decrease order by order in loops.
Third, that natural sets of parameters exist at the three-loop level; note that natural parameter 
sets also exist in the full Brueckner calculation \cite{ref:Hu00}. 
Fourth, this is not a Brueckner calculation, so there was no guarantee the three-loop terms would be small. Fifth,
the local contributions appear to be contained in the parameterization of QHD.
To fully decide whether or not the loop expansion is asymptotic for QHD, one needs to consider the ladder
and ring sums.


\section*{Acknowledgments}
We would like to thank Dr.\ B. D. Serot for his useful comments during the course of this
work and on the manuscript. This work was supported in part by the Department
of Energy under Contract No.\ DE--FG02--87ER40365.

%

% \appendix

% \section{}
% \label{}

\end{document}